\documentclass{aa}
\usepackage{graphicx}
\usepackage{longtable}
\usepackage{txfonts}
\usepackage{natbib}
\bibpunct{(}{)}{;}{a}{}{,} 

\newcommand\T{\rule{0pt}{2.0ex}}       

\newcommand{\D}{{\mathrm d}}

\newcommand{\SBrak}[1]{{\left[{#1}\right]}}

\newcommand{\DxDyInd}[3]{{\Brak{\frac{\D{#1}}{\D{#2}}}_{{#3}}}}
\newcommand{\dxdycz}[3]{{\Brak{\frac{\partial{#1}}{\partial{#2}}}_{{#3}}}}
\newcommand{\Brak}[1]{{\left({#1}\right)}}

 \def\mso{\,\mathrm{M}_\odot}
 \def\rso{\,{\rm R}_\odot}
 \def\lso{\,{\rm L}_\odot}
 
 \def\kms{\, {\rm km}\, {\rm s}^{-1}}
 
 \def\teff{\log\, T_{\rm }\,}
 \def\llso{\log\, L/ L_\odot \,}
 \def\simle{\mathrel{\hbox{\rlap{\hbox{\lower4pt\hbox{$\sim$}}}\hbox{$<$}}}}
 \def\simgr{\mathrel{\hbox{\rlap{\hbox{\lower4pt\hbox{$\sim$}}}\hbox{$>$}}}}

 \def\grad{\nabla}
 \def\adrad{\nabla_{\mathrm{\!rad}}}
 \def\adgrad{\nabla_{\mathrm{\!ad}}}
 \def\mugrad{\nabla_{\!\mu}}
 
 \def\cp{c_{\!P}}
 \def\hp{\mathrm{H}_{\mathrm{P}}}
 \def\hpc{\mathrm{H}_{\mathrm{P,c}}}
 \def\hps{\mathrm{H}_{\mathrm{P,s}}}
 \def\c2{^{12}{\rm C}}
 \def\c3{^{13}{\rm C}}
 \def\n14{^{14}{\rm N}}
 \def\c1213{^{12}{\rm C}/^{13}{\rm C}}
 \def\he3he4{^3{\rm He}/^4{\rm He}}
 \def\Ra{\mathrm{Ra}}
 \def\logt{\log\, T}
 
 \def\microt{\xi}
 \def\vsini{\varv\sin i}
 \def\vc{\varv_{c}}
 \def\vca{\langle{\varv_{c}}\rangle}
 
 \def\Rtop{\mathrm{R}_{\mathrm{c}}}
 
 \def\flt{\tau_{fl}}
 \def\rst{\,{\rm R}_\star}
 \def\vin{\varv_{\,\mathrm{\infty}}}
 \def\cp{c_{\mathrm{p}}}
 \def\ttwenty{6.41$\times10^6$ }
 \def\tsixty{2.37$\times10^6$ }

\begin{document}
   \title{Sub-surface convection zones in hot massive stars and
 their observable consequences}
   
   \author{M. Cantiello\inst{1},
           N. Langer\inst{1,2},
           I. Brott\inst{1},
           A. de Koter\inst{1,3},
           S. N. Shore\inst{4},
           J. S. Vink\inst{5},
           A. Voegler\inst{1},
           D. J. Lennon\inst{6}
           \and
           S.-C.~Yoon\inst{7}}
   \offprints{M. Cantiello \email{m.cantiello@uu.nl}}

   \institute{Astronomical Institute, Utrecht University,
              Princetonplein 5, 3584 CC, Utrecht, The Netherlands
             \and          
             Argelander-Institut f\"ur Astronomie der Universit\"at Bonn, Auf dem H\"ugel 71, 53121 Bonn, Germany 
             \and
              Astronomical Institute Anton Pannekoek, University of Amsterdam, Kruislaan 403, 1098 SJ, Amsterdam, The Netherlands
             \and
              Dipartmento di Fisica ``Enrico Fermi'', Universit\`a di Pisa, via Buonarroti 2, Pisa 56127 and INFN - Sezione di Pisa, Italy
             \and
              Armagh Observatory, College Hill, Armagh, BT61 9DG, Northern Ireland (UK)
             \and
             Space Telescope Science Institute, 3700 San Martin Drive, Baltimore, MD 21218, USA  
             \and
               Department of Astronomy \& Astrophysics, University of California, Santa Cruz, High Street, Santa Cruz, CA 95064, USA
              }

   \date{Received 12 January 2009 / Accepted 03 March 2009}

\abstract
    {We study the convection zones in the outer envelope of hot massive stars which are 
   caused by opacity peaks associated with iron and helium ionization.}
  {We determine the occurrence and properties of these convection zones as function of the stellar parameters. 
   We then confront our results with observations of OB stars.} 
   {A stellar evolution code is used to compute a grid of massive star
   models at different metallicities. In these models, the mixing length theory is used to
   characterize the envelope convection zones.} 
   {We find the iron convection zone (FeCZ) to be more prominent for lower surface gravity,
   higher luminosity and higher initial metallicity.  
   It is absent for luminosities below about $10^{3.2}\lso$, $10^{3.9}\lso$, 
   and $10^{4.2}\lso$ for the Galaxy, LMC and SMC, respectively.
   We map the strength of the FeCZ on the Hertzsprung-Russell 
   diagram for three metallicities, and compare this with the occurrence of
   observational phenomena in O stars: microturbulence, non-radial pulsations, 
   wind clumping, and line profile variability.}
   {The confirmation of all three trends for the FeCZ
   as function of stellar parameters by empirical microturbulent velocities
   argues for a physical connection between sub-photospheric convective
   motions and small scale stochastic velocities in the photosphere of
   O- and B-type stars.
   We further suggest that clumping in the inner parts of the winds of OB stars
   could be caused by the same mechanism, and that magnetic fields produced in the
   FeCZ could appear at the surface of OB stars as diagnosed
   by discrete absorption components in ultraviolet absorption lines.}

   \authorrunning{M. Cantiello et al.}
   \titlerunning{Sub-surface convection in hot stars}
   \maketitle


\section{Introduction}

Massive stars, in a general sense, have convective cores and radiative
envelopes \citep{Kw90}.  The introduction 
of the so called ``iron peak" in stellar opacities \citep{Irw92} led, however,  
to the prediction of  a small convection
zone in the envelope of sufficiently luminous massive main sequence models
\citep{Sc93}.  It is often accompanied by an even smaller convection zone
which originates from an opacity peak associated with 
partial helium ionization.  These two convection zones comprise 
almost negligible amount of mass.  
The reality of the iron opacity bump, as predicted by various groups \citep[e.g.,][]{Irw92,2005MNRAS.360..458B},
is unambiguous. It is most obvious in the field of
stellar pulsations. Only the inclusion of this feature allows an agreement of
observed and predicted instability regimes in the HR diagram, from the white
dwarf regime \citep[e.g.][]{1993MNRAS.260..465S,1997ApJ...483L.123C}, for main sequence stars \citep[e.g., $\beta~$Cephei stars; see]
[and references therein]{2001MNRAS.327..881D}, and up to hot supergiants \citep{2006ApJ...650.1111S}.

While the envelope convection zones may, at first glance, be negligible for the internal 
evolution of hot massive stars, they may cause observable
phenomena at the stellar surface. The reason is that
the zones are located very close to the photosphere for some mass 
interval (see below).
Here, we will discuss which observed features in hot stars might be
produced by these near surface convection zones. In particular, we examine 
whether a link exists between these convective regions and 
observable small scale velocity fields at the stellar surface 
and in the stellar wind, ``microturbulence". 
A similar idea has been used to explain microturbulence 
in low mass stars \citep{Edm78}, in which deeper 
envelope convection zones reach the photosphere. 
While \citet{Edm78} concludes that the 
same mechanism {\em cannot} explain microturbulent velocities in O and B stars, 
the iron-peak induced sub-photospheric convection zones in these stars had not yet been 
discovered.  We demonstrate in this paper that these convection zones may not only
cause motions which are observable, but possibly even directly affect the evolution:
First, we discuss how photospheric velocity fields may affect the 
structure of massive star winds by inducing clumping at the base of the wind and thereby affecting the 
stellar mass-loss. And second, we argue that the near surface convection zones
may generate magnetic fields which -- if they migrate to the surface --  further affect
the stellar wind mass-loss and,  more significantly, the associated stellar angular momentum
loss.

We construct grids of massive main sequence star models, for various metallicities, 
that allow us to predict the occurrence and properties of sub-surface convection zones as function
of the stellar parameters (Sect.~\ref{results}). We then compare the model predictions with
observed stellar properties, e.g., empirically derived microturbulent velocities 
and observations of wind clumping in hot massive stars (Sect.~\ref{comparison}). 

\section{Method}\label{method}
Our stellar models are calculated with a hydrodynamic stellar evolution code.
This code can calculate the effect of rotation on the stellar structure, rotationally induced
chemical mixing, and the transport of angular momentum by magnetic torques \citep[see][and references therein]{ply+05,yln07}.
Compositional mixing is treated as a diffusive process. The rate of change of a nuclear 
species of mass fraction $X_i$ is calculated as
\begin{equation}
  \dxdycz{X_i}{t}{}=\dxdycz{}{m}{}\, \SBrak{(4\pi r^2 \rho)^2 \, D \,
     \dxdycz{X_i}{m}{}}+\DxDyInd{X_i}{t}{\mathrm{nuc}} ,
\end{equation}
where $D$ is the diffusion coefficient constructed from the sum of
individual diffusion coefficients for the range of mixing processes \citep[see][and references therein]{hlw00}.
The second term on the right hand side is the schematic symbol to stand for 
all nuclear reactions. The contributions 
to the diffusion coefficient are convection, semiconvection and
thermohaline mixing. For rotating models also the contributions from rotationally induced mixing 
and magnetic diffusion are computed. The transport of angular momentum is also treated 
as a diffusive process \citep{es78,pin89,hlw00}. 

The Ledoux criterion is used to determine 
which regions of the star are unstable to convection:
\begin{equation}
\adgrad - \grad +\frac{\varphi}{\delta} \,\mugrad \le 0
\end{equation}
\citep[e.g.,][]{Kw90} where $\adgrad$ is the adiabatic temperature gradient and $\mugrad$ is the gradient in the mean molecular
weight. 
The diffusion coefficient, $D$, in convective regions is approximated with 
\begin{equation}
D=\frac{1}{3} \alpha \hp \varv_c
\end{equation}
where $\hp$ is the pressure scale height, $\varv_c$ is the convective velocity,  
and $\alpha$ the mixing length parameter.  We fix $\alpha = 1.5$, which results from 
evolutionary tracks of the Sun 
\citep[e.g.][]{1997ApJ...480..395A,1999A&A...346..111L}; a sensitivity study of 
the $\alpha$ dependence of our scenario will be presented in future work. 
The convective velocity, $\varv_c$, is calculated using the
mixing length theory \citep{BV58} (MLT hereafter) and the convective contribution to the diffusion 
coefficient becomes:
\begin{equation}
D=\frac{1}{3}\,\alpha^{2/3} \hp \,\Big[\frac{c}{\varkappa \rho}\, g\, \beta\,(1-\beta) \adgrad(\adrad-\adgrad)  \Big]^{1/3}  ,
\end{equation}
where $\varkappa$ is the opacity, $\rho$ is 
the density, $\beta$ is the ratio of gas pressure to total pressure, 
$g$ is the local gravitational acceleration, and $c$ is the speed of light. 
Here, $\adrad$ and $\adgrad$ are the radiative and adiabatic gradients, respectively. 

\begin{figure}
  \resizebox{\hsize}{!}{\includegraphics{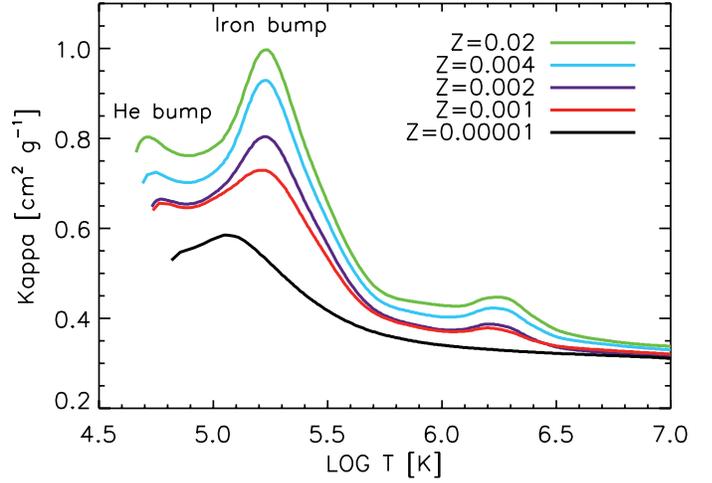}}
  \caption{Opacity in the interior of  60 $\mso$ zero age main sequence stars
of various metallicities (see legend) as a function of temperature, from the surface
up to a temperature of $10^7\,$K.  
   The different colors refer to different metallicities, as shown in the legend. 
}
  \label{varkappa}
\end{figure}

We use the solar composition proposed by  \citet{2005ASPC..336...25A}. The opacities in our code are extracted from the OPAL tables \citep{ir96}. 
Fig.~\ref{varkappa} shows the opacity coefficient as 
function of temperature in our 60$\mso$ models 
for various metallicities. The peaks at $\logt \simeq 4.7$ and $\logt \simeq 5.3$ 
are caused by helium and iron, respectively. The peak at  $\logt \simeq 6.2-6.3$ 
is caused by carbon, oxygen and iron. 

We use the metallicity dependent mass-loss predictions of \citet{2001A&A...369..574V}.

\subsection{The helium convection zone}\label{he}
In the very weak helium convection zone, radiative diffusion is 
the dominant energy transport mechanism, 
which may have consequences for the development of convection. 
In fact, in viscous fluids the Ledoux-criterion is not
strictly correct, since it ignores any dissipative 
effect on the evolution of a perturbation.
A more accurate criterion can be expressed in terms of the non-dimensional Rayleigh
number, $\Ra$ which for compressible, stratified convection, is
\begin{equation}
\Ra \simeq \frac{(\grad-\adgrad)  L^3  g}{  \kappa\nu} .
\label{rayleigh}
\end{equation}
Here $L$ is the thickness of the convective layer, and $\kappa$ 
and $\nu$ are, respectively, 
the thermal diffusivity and the kinematic (molecular) 
viscosity \citep[e.g,][p. 328]{1992aita.book.....S}.

For convection to develop, $\Ra$ must exceed some critical value, $\Ra_c$. 
The estimate of $\Ra$ in the helium convective 
region depends on the choice of the viscosity coefficient. For the Spitzer formula 
\citep{1962pfig.book.....S},$\Ra > \Ra_c$, and the region 
can be considered convective. In contrast, for the radiative viscosity 
\citep[e.g,][p. 445]{Kw90}, $\Ra < \Ra_c$.  
There is an additional uncertainty in these estimates since the expressions 
for the radiative transport coefficients in our models
are strictly correct only in the diffusion limit. 
Likewise, the value of the heat capacity $\cp$ 
can vary by
an order of magnitude depending on whether the radiative energy reservoir $a T^4$ 
is coupled to the internal energy of the gas or not. Since the helium convection zone
occurs very close to the surface in our models, these 
additional uncertainties could be relevant.

Ideally, the properties of the helium convection zone could be studied
through multi-dimensional hydrodynamic calculations. However,
the large thermal diffusivity poses a 
formidable computational challenge since it makes the
problem numerically stiff: the diffusive timescale is much shorter than the dynamical one, which leads
to very short time steps if an explicit solver is used (unfortunately, most codes used for compressible
convection are explicit). Any simulation would have only limited value unless it includes a
sufficiently realistic treatment of the coupling between plasma and radiation.

In the presence of strong wind mass-loss, 
another consideration related to the He convective zone becomes important,
due to the fact that it comprises only a tiny amount of mass.  
Convection can set in only if the turnover time  
$\tau_{\rm turn}\simeq \hp/\varv_c$ is shorter than the time scale for
which convection is predicted to prevail at a fixed Lagrangian mass shell
inside the convection zone, $\tau_{\rm conv}$, which is   
$\tau_{\rm conv} \simeq \Delta M_{\rm conv}/\dot M$. 
We find a critical mass-loss rate $\dot{M} \sim 10^{-6} \mso \rm{yr}^{-1}$, above which 
convection has no time to develop in the helium region, since the wind is removing
an amount of mass equivalent to the mass of the convection 
zone before a convective eddy can turn over (see Tab.~\ref{clumps}).
For a metallicity $Z$=0.02, stars above 40 $\mso$ cannot
develop the He convection zone, and in a 20 $\mso$ such a layer is
convective only  for  10 - 100 turnovers before convection moves to a lower
mass coordinate.  
None of these concerns is significant for the iron convection zone (FeCZ hereafter),
where convection is always fully developed.
Moreover the convective velocities for the FeCZ are always
found to be much higher than those in the helium convection zones.
We disregard the occurrence of the helium convection zones unless it
is explicitly mentioned. 

\section{Results}\label{results}

We calculated a grid of non-rotating stellar evolution sequences for initial masses
between $5\mso$ and $100\mso$, at metallicities of  $Z$=0.02,  $Z$=0.008 and  $Z$=0.004, roughly corresponding to
the Galaxy, the LMC and the SMC, respectively. Additionally, we computed several models at
lower metallicity. Since rapid rotation can change the properties of
sub-surface convection \citep{2008A&A...479L..37M}, we calculated a few rotating models to evaluate the effects of rotation on our results. 
These effects are discussed in Section~\ref{rotation}.

\begin{figure}
  \resizebox{\hsize}{!}{\includegraphics{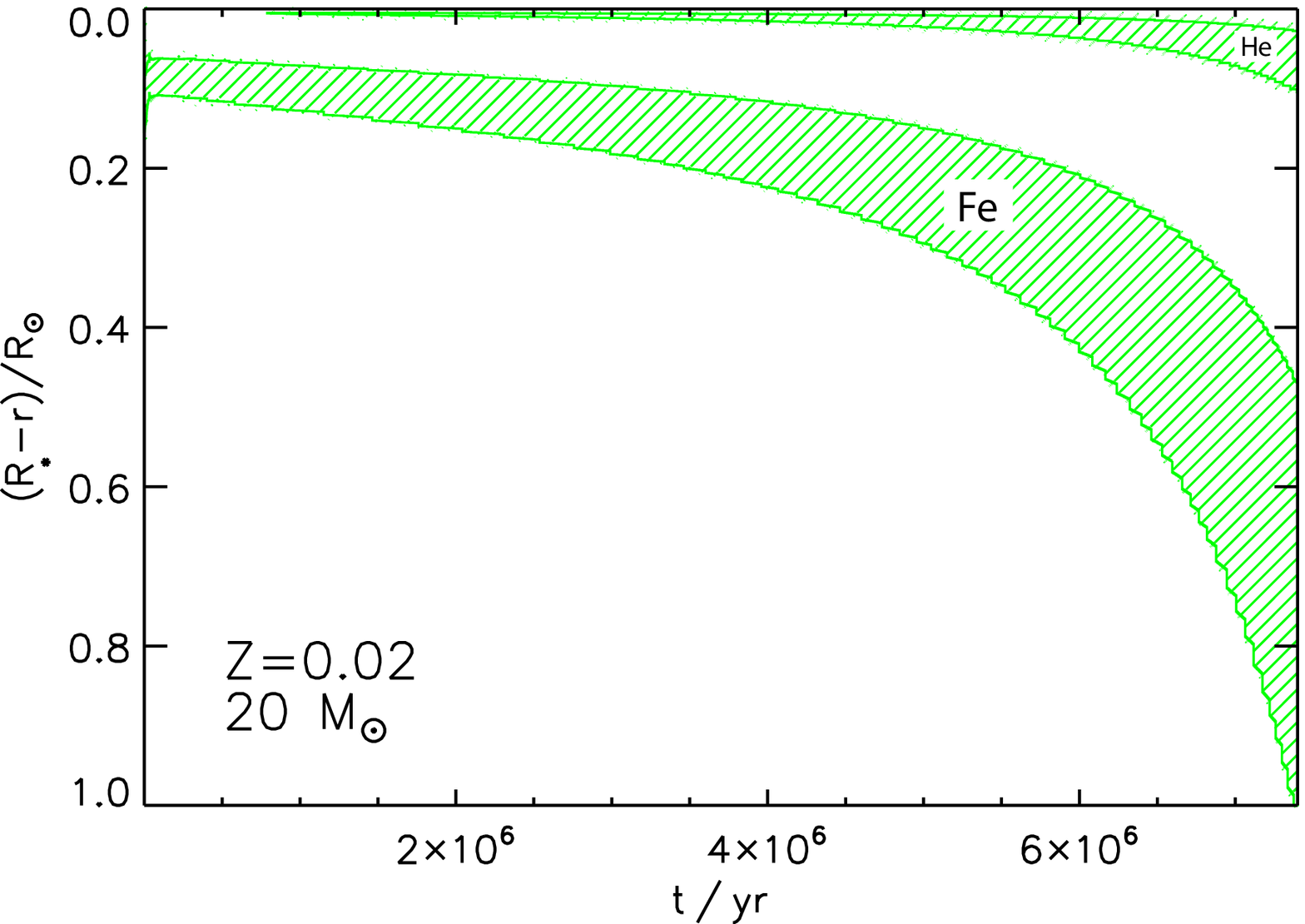}}
  \resizebox{\hsize}{!}{\includegraphics{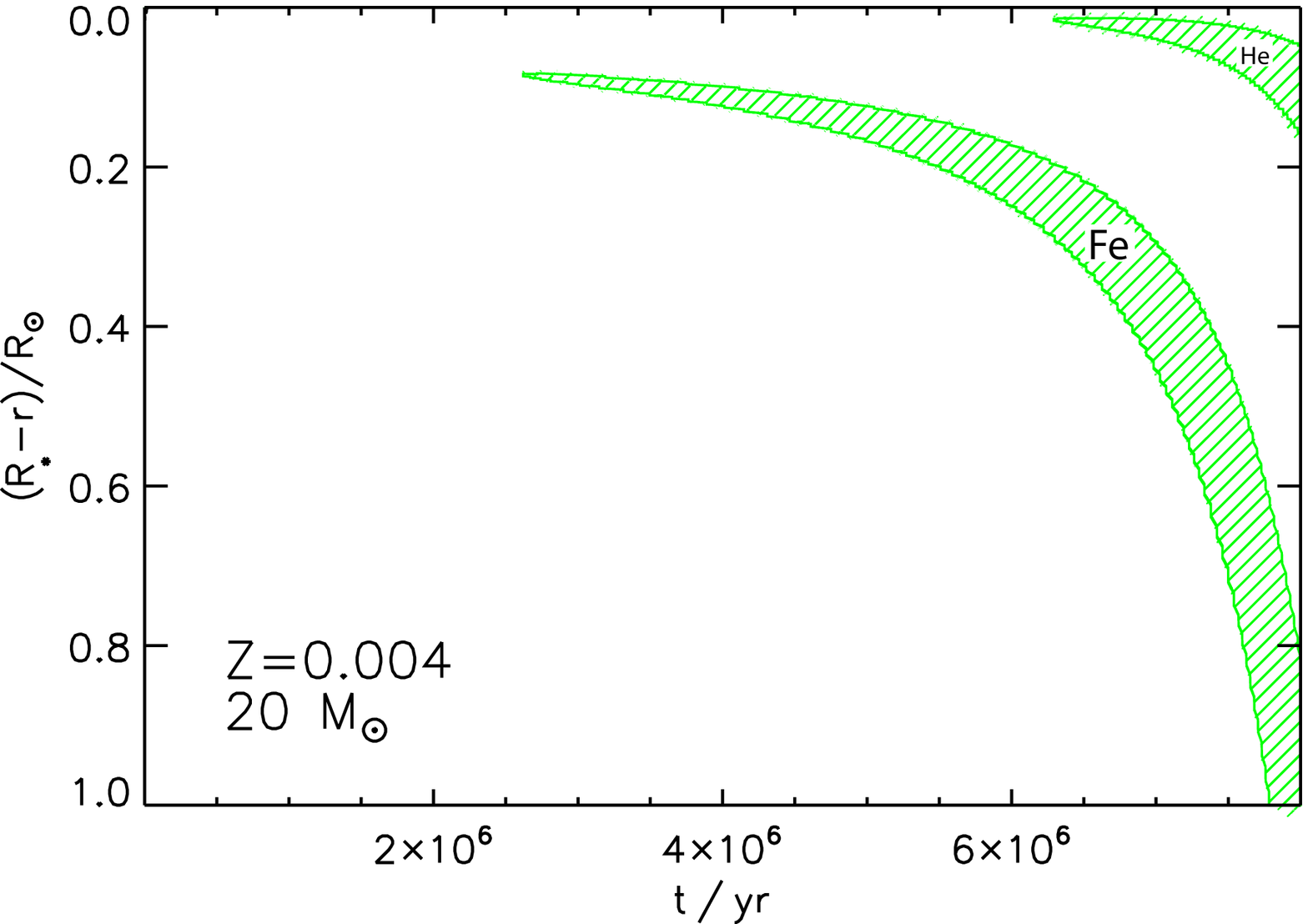}}
  \caption{Evolution of the radial extent of the helium and iron convective regions 
   (hatched) as function of time, 
   from the zero age main sequence to roughly the end of core hydrogen burning, for a 20$\mso$ star. The top of the plot
   represents the stellar surface. Only the upper 1$\rso$ of the star is shown in the plot, 
   while the stellar radius itself increases during the evolution. 
   Upper panel: The star has a metallicity of $Z$=0.02, and its effective temperature 
   decreases from 35\,000\,K to 25\,000\,K during the main sequence phase.
   Lower panel: The star has a metallicity of $Z$=0.004, and its effective temperature
   decreases from 37\,000\,K to 27\,000\,K during the main sequence phase. The extent of the 
   convection zones is smaller than in the case shown above, and the iron zone is absent for the
   first 2.5~million years. } 
  \label{20zsunsurf}
\end{figure}

\begin{figure}
 \resizebox{\hsize}{!}{\includegraphics{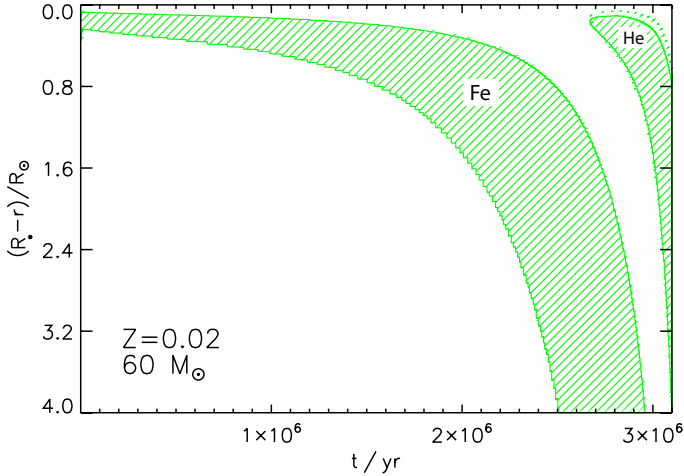}}
  \caption{Same as Fig.~\ref{20zsunsurf}, but for a 60$\mso$ star at $Z$=0.02. Note the different vertical scale, spanning the upper 4$\rso$ of the star. 
   The effective temperature decreases from 48\,000\,K to 18\,000\,K during the main sequence phase.}
  \label{60zsunsurf}
\end{figure}

Figures~\ref{20zsunsurf} and~\ref{60zsunsurf} show the evolution 
of the radial extent and location of the sub-surface convection zones in
20$\mso$ and 60$\mso$ models during the main sequence phase.

As outlined above, 
the He opacity bump at around $\logt~\simeq~4.7$ is responsible for a
convective zone which occurs close to the stellar surface and is very inefficient: 
only a very small fraction of the heat flux is transported by bulk motions 
in this region.
The upper boundary is typically found at an optical depth in the range 
$2 \leq \tau \leq 10$, where $\tau$ is the Rosseland mean optical depth.  
Below this convective zone, the Fe opacity bump at around $\logt~\simeq~5.3$ is 
associated with a more efficient and extended convective region.

The radial extent of the FeCZ is quite substantial,
i.e. a significant fraction of one solar radius, which corresponds
typically to 2 - 10 pressure scale heights, comprising a 
mass on the order of 10$^{-6}\mso$ to 10$^{-5}\mso$, while 
the amount of mass between the top of the FeCZ and the stellar
surface is around several times 10$^{-7}\mso$ (cf. Table~1). 
In the 20$\mso$ model the upper border of the FeCZ 
is located at $\tau \approx 140$ on the ZAMS, and at 
$\tau \approx $ 370 on the cool side of the main sequence band. In 
the 60$\mso$ model the upper border at ZAMS is located at 
$\tau \approx $ 15, reaching $\tau \approx $ 260 during 
the late main sequence evolution. 
Convective velocities predicted by the MLT are on the order of 10s of $\kms$,
where more extended zones achieve higher velocities.
For a quantitative analysis, we define an average convective velocity 
\begin{equation}
\vca:=\frac{1}{\alpha \hp} \int_{\Rtop-\alpha\hp}^{\Rtop} \vc(r)\; dr
\label{aver}
\end{equation}
where $\Rtop$ is the upper boundary of the 
convective zone, and where we set $\alpha = 1.5$.

\begin{table*}

\begin{minipage}[t]{\textwidth}
\caption{Properties of the envelope convection zones in our 20 and 60 $\mso$ models of solar metallicity. These are the same models shown in the top panel of Fig.~\ref{20zsunsurf} and in Fig.~\ref{60zsunsurf}. The values in the table refer
to t=\ttwenty  for the 20 $\mso$ model and t=\tsixty for the 60 $\mso$ model.}
\label{clumps}
\renewcommand{\footnoterule}{}
\centering
\begin{tabular}{l | c c c c c c c c c c}
\hline\hline
   M        \T  & Zone    &      $\hp$ & $\vca$  & $\Delta M_{\rm conv}$\footnote{Mass contained in the convective region.}  & $\Delta M_{\rm top}$\footnote{Mass in the radiative layer between the stellar surface and the upper boundary of the convective zone.}  & $N_{\rm cells}$\footnote{Expected number of convective cells, $N_{\rm cells}:= (R_{\star}/H_{\rm P})^2$.}  & $\tau_{\rm turn}$\footnote{Convective turnover time, $\tau_{\rm turn}:= H_{\rm P}/\vca$.} &   $\tau_{\rm conv}$\footnote{Time that a piece of stellar material spends 
inside a convective region, $\tau_{\rm conv}:= \Delta M_{\rm conv}/\dot M$.} & $ \dot{M}$ \\
 M$_{\sun}$ \T  &         & R$_{\sun}$ &  km~s$^{-1}$              & M$_{\sun}$             & M$_{\sun}$            &                  &       days          &  days     & $\mso \rm{ yr}^{-1}$ \\

\hline
20              & He      & 0.025      &   0.08       & $7.6\times10^{-9}$     &  $1.9\times10^{-9}$     & $1.8\times10^{5}$ &   2.5        &  38         &      $7.3\times10^{-8} $      \\
20              & Fe      & 0.08       &   2.40       & $3.6\times10^{-6}$     &  $5.8\times10^{-7}$     & $1.8\times10^{4}$  &   0.25       &  18250  &  $7.3\times10^{-8}$   \\
60              & Fe      & 0.24       &   2.25       & $1.6\times10^{-5}$     &  $9.8\times10^{-7}$     & $8.5\times10^{3}$  &   0.83       &   1570    &  $3.7\times10^{-6}$      \\
\hline

\end{tabular}
\end{minipage}
\end{table*}

From Figures~\ref{20zsunsurf} and~\ref{60zsunsurf}, three trends for the extent 
of the sub-surface convection zones are noticeable.  First, with increasing 
time during the main sequence evolution, 
these zones become more extended, and are
located deeper inside the stellar envelope. This is because the stellar envelope expands,
and becomes cooler, while the temperature of the opacity peak remains nearly constant.
In our $20\mso$ model at $Z$=0.02, the mass of the He convective zone increases  
from about $10^{-9}\mso$ to $2\times10^{-7}\mso$, and that of the FeCZ is growing from 
$2\times10^{-6}\mso$ to $10^{-4}\mso$.
For sufficiently hot models, the helium convection zones can even vanish (Fig.~\ref{20zsunsurf}, lower panel).
Second, comparing the 20$\mso$ and the 60$\mso$ model at $Z$=0.02 demonstrates that the
FeCZ becomes more prominent for higher luminosity. This is because the opacity is not substantially changing 
among main sequence models at the same metallicity, such that a higher luminosity 
renders a larger portion of the envelope convectively unstable (both in radius and mass fraction).   
Our models show that the FeCZ disappears below a threshold luminosity
of about $10^4\lso$ on the ZAMS at solar metallicity. Third, comparing the two
20$\mso$ models in Fig.~\ref{20zsunsurf} shows that the extent of the FeCZ,
and its presence, depends on the metallicity. We find that for $Z$=0.001, it 
is completely absent below 40$\mso$, and at $Z$=0.00001 it does not occur for $\mathrm{M}\leq60\mso$.
In summary, our models predict an increase of the importance of the FeCZ for 
cooler surface temperature or lower surface gravity, for higher luminosity, and for higher metallicity. 

While in the discussed range of luminosity and effective temperature, 
the average convective velocity $\vca$  is on the order of 1 to 10$\kms$ for the 
FeCZ, 
we found that the average convective velocity $\vca$
in the He convective zone is always very low ($\simle 1\kms$). 
Convection due to hydrogen
recombination is absent; this dominates at lower effective temperatures 
than the ones studied here.

For our grid of stellar evolution models, 
we map the average convective velocity  of the FeCZ (Eq.~\ref{aver}) in the 
HR diagram for the three different metallicities (see Fig.~\ref{hrdvel}, and 
Sect.~\ref{mobs}).  This figure displays the three qualitative 
trends of the iron zone we have just described.

\begin{itemize}
\item For given luminosity and metallicity, the average convective velocity near the
upper boundary of the FeCZs increases with decreasing surface temperature.
The convection zones are located deeper inside the star (in radius, not in mass), 
and the resulting larger pressure scale height leads to higher velocities. 
At solar metallicity and $10^5\lso$ (i.e. roughly at $20\mso$) the velocities
increase from just a few $\kms$ at the ZAMS to more than $10\kms$ in the supergiant regime,
where $\vca = 2.5\kms$ is achieved at $T_{\rm eff}\simeq 30\, 000\,$K.
At the lowest considered metallicity, the FeCZ
is absent at the ZAMS at $10^5\lso$, and a level of $\vca = 2.5\kms$ is only reached
at $T_{\rm eff}\simeq 20\, 000\,$K.

\item For fixed effective temperature and metallicity, the 
iron zone convective velocity
increases with increasing luminosity, since a larger flux 
demanded to be convectively transported 
requires faster convective motions. Figure \ref{hrdvel} in Sect.~\ref{mobs} also 
shows that there are
threshold luminosities below which FeCZs do not occur, i.e.,
below about $10^{3.2}\lso$, $10^{3.9}\lso$, and $10^{4.2}\lso$ for the Galaxy, 
LMC and SMC, respectively.

\item The FeCZs become weaker for lower metallicities,
since due to the lower opacity, more of the flux can be transported by radiation. 
The threshold luminosity for the occurrence of the FeCZ 
quoted above for $Z$=0.02 is ten times lower than that
for $Z$=0.004. And above the threshold, for a given point in the HR diagram, the convective velocities
are always higher for higher metallicity. 
\end{itemize}

\subsection{Rotating models}\label{rotation}
\begin{figure}
 \resizebox{\hsize}{!}{\includegraphics{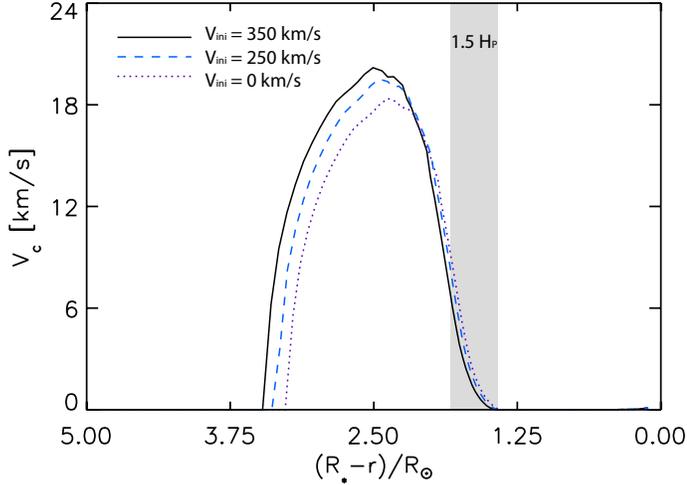}}
  \caption{Convective velocity in the FeCZ as function of radial distance from the stellar surface. The dotted line  
  corresponds to a non-rotating 20$\mso$ model at $Z$=0.02, while the dashed and solid lines refer to the same model rotating at birth with 250$\kms$ and 350 $\kms$ 
  respectively. The values correspond to models having the same effective temperature ($\teff=$ 4.339) and very similar luminosity ($\llso=$ 5.04 for the non-rotating model and
  $\llso=$ 5.03 for the rotating ones). The gray band shows the upper 1.5 pressure scale heights of the FeCZ, which is the region considered for the computation of $\vca$, cf. Eq.~\ref{aver}. 
  Convective velocities in the He convection zone are much lower than 1$\kms$ and are not visible in this plot.}
  \label{feczrotating}
\end{figure}

 We considered two 20$\mso$ models with metallicity $Z$=0.02, one rotating at birth with an equatorial velocity 
of 250$\kms$ (corresponding to about 35\% of the critical velocity) and one with 350$\kms$
 (about 50\% of the critical velocity). 
The evolution of the radial extent of sub-surface convection in the rotating models is very similar to the non-rotating case shown in Fig.~\ref{20zsunsurf}. 
Also the convective velocities inside the FeCZ change only a few percent between rotating and non-rotating models, 
even if the rotating models show slightly higher convective velocity peaks (see Fig.~\ref{feczrotating}). We conclude that rotation is not significantly
 affecting the structure and the properties of sub-surface convection in the vast majority of OB stars.

As pointed out by \citet{2008A&A...479L..37M}, the effects of rotation on sub-surface convection become substantial for stars rotating close to critical velocity.  While stars rotating with such high velocities 
exist (e.g. Be stars), their number is modest. The study of sub-surface convection in these very fast rotators is interesting, but may require 2-dimensional stellar models, which is beyond the scope of this paper.

\section{Comparison with observations}\label{comparison}

In the following, we investigate the idea that these sub-surface convection zones 
might be related to observable phenomena at the stellar surface.
In particular, we investigate potential connections with microturbulence in massive stars,
and discuss whether small scale or large scale clumping in massive star winds, 
magnetic fields, and non-radial pulsations could be related to sub-surface convection. 
For each point, we first briefly discuss the theoretical motivation, and then the corresponding 
observational evidence.

\subsection{Microturbulence}\label{micro}

\subsubsection{Theoretical considerations}\label{waves}
The convective cells in the upper part of a convection zone excite 
acoustic and gravity waves that propagate outward. 
The generation of sound waves by turbulent motions was first discussed 
by \citet{Lig52} and extended to a stratified atmosphere by  \citet{Ste67} and \citet{gk90}.
In a stratified medium, gravity acts as a restoring force and allows the excitation
of gravity waves. For both acoustic and gravity waves, the most important parameter
determining the emitted kinetic energy flux is the velocity of the convective motions.
This is why, in the following, we use the average convective velocity $\vca$
as the crucial parameter determining the efficiency of sub-surface convection.

\begin{figure}
\resizebox{\hsize}{!}{\includegraphics{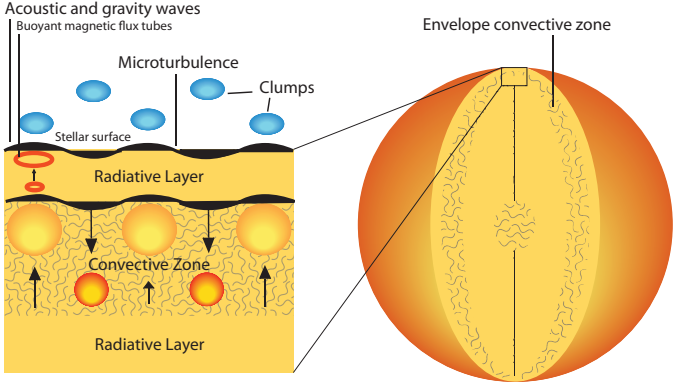}}
  \caption{Schematic representation of the physical processes 
connected to sub-surface convection. Acoustic and gravity waves emitted in the convective zone
   travel through the radiative layer and reach the surface, inducing density and velocity fluctuations. 
In this picture microturbulence and clumping at the base of the wind are a
   consequence of the presence of sub-surface convection.  Buoyant magnetic flux tubes produced
   in the convection zone could rise to the stellar surface.}
  \label{sketch}
\end{figure}

\citet{gk90} showed that convection excites acoustic and gravity waves,
resulting in maximum emission for those waves with horizontal
wave vector $k_h\sim 1/\hpc$ and angular frequency $\omega\sim\varv_c/\hpc$, 
where now $\varv_c$ and $\hpc$ are evaluated at the top of 
the convective region. They calculated that 
the amount of convective kinetic energy flux
going into acoustic and gravity waves is
\begin{equation}
F_{{\mathrm ac}}\sim F_{c} M_c^{15/2} ,
\end{equation}
and
\begin{equation}
F_{{\mathrm g}}\sim F_{c} M_c  ,
\end{equation}
respectively,
where we take $F_{c}\sim \rho_c \vca^3$ and $M_c$ is the Mach number in 
the upper part of the convective region.
Since convection in our models is subsonic, gravity waves are expected 
to extract more energy from the convective region than acoustic waves.
These gravity waves can then propagate outward,
reach the surface and induce observable density and velocity fluctuations (Fig.~\ref{sketch}).

The Brunt-Vais\"{a}la frequency in the 
radiative layer above the FeCZ is about mHz.
Molecular viscosity can only damp the highest frequencies,  while wavelengths
that will be resonant with the scale length of the line forming region should
not be affected \citep[see e.g.][]{Lig67}. This is the case for the gravity waves
stochastically excited by convective motions: they can easily propagate through
the sub-surface radiative layer, steepening and becoming dissipative only in
the region of line formation.

Again, multi-dimensional hydrodynamic simulations would be the best way to 
compute the energy loss of these waves during their propagation through the
radiatively stable envelope above the FeCZ, but this is
beyond what we can presently do.  We can, however, obtain 
an upper limit to the expected velocity amplitudes at the stellar surface,
where we only consider the energy transport through gravity waves.
The kinetic energy per unit volume associated with the surface velocity
fluctuations $E_{\rm s}$ must be comparable to or lower than the
kinetic energy density associated 
with the waves near the sub-surface convection zone, $E_{\mathrm
  g}\sim M_c\,\rho_c\,\vca^2 $, or
\begin{equation}
\frac{E_{{\mathrm g}}}{E_{\rm s}}\sim M_c\left( \frac{\rho_c}{\rho_s} \right)\left(\frac{\vca}{\varv_{\rm s}}\right)^2\geq 1   ,
\label{ratio}
\end{equation}
where $\rho_c$ is the density at the top of the convective region and $\rho_s$ is the surface density,
and $\varv_{\rm s}$ is the surface velocity amplitude. In this ratio
we only consider energy density since the
volume of the line forming region is comparable to the volume of the
upper part of the convective zone.  Therefore, we expect
\begin{equation}
\varv_{\rm s} \leq \vca \sqrt{M_c {\rho_c \over \rho_s}} 
\end{equation}
In our models with well developed FeCZs,  
$\sqrt{M_c \,{\rho_c / \rho_s}} \simeq 1$ (order of magnitude), and thus
$\varv_{\rm s}$ and $\vca$ should be on the same order of magnitude. 
It is difficult to estimate the typical correlation length 
of the induced velocity field at the stellar surface, but a 
plausible assumption is that it is about one 
photospheric pressure scale height, $\hps$, 
given the proximity of the FeCZ to the surface 
and the fact that the horizontal wave vector of the 
emitted waves is $k_h\sim1/\hpc$.

\subsubsection{Observations}\label{mobs}

The microturbulent velocity  $\microt$ is defined as the microscale nonthermal component of the
gas velocity in the region of spectral line formation:  
\begin{equation}
\Delta\lambda_D = \frac{\lambda}{c}\,\sqrt{\frac{2 \mathcal{R} T}{\mu} + \microt^2}
\end{equation}\label{microturbulence}
Assuming that the gas in this zone has 
a temperature only slightly different from the effective temperature,
one finds empirically that the observed Doppler widths $\Delta\lambda_D$ 
cannot be accounted for by the thermal
motions alone \citep[e.g.][]{Cow70}.
Regardless of which physical mechanism causes microturbulence, the process of spectral line
fitting yields values of $\microt$ in hot massive stars between 0 and about 20$\kms$.  
In contrast, macroturbulence corresponds to velocity fluctuations which are coherent
on a length scale larger than the radial extent of line forming regions. If indeed 
the length scale of the photospheric velocity fluctuations induced by the iron convection 
zone are on the order of the photospheric pressure scale height, then this length scale is
also comparable to the radial extent of line forming regions, and it is difficult to decide
whether the velocity fluctuations would be manifested as micro- or as macroturbulence, or both.
Below, we compare our model predictions only to the case of microturbulence since this 
is the empirical parameter most extensively available in the literature.

\begin{figure}
  \resizebox{\hsize}{!}{\includegraphics{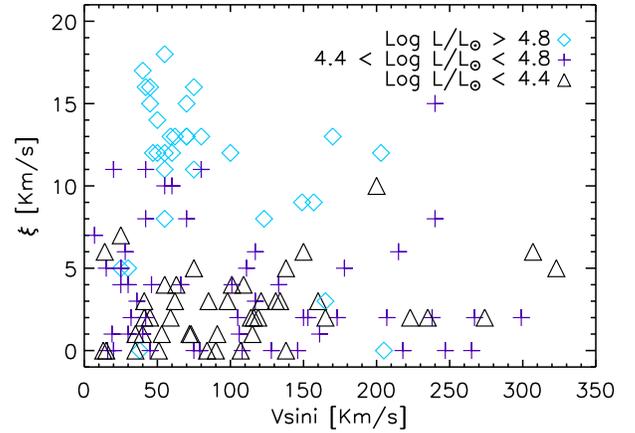}}
  \caption{Projected rotational velocity $\vsini$ versus photospheric microturbulent
velocity $\microt$ for the early B-type stars in the LMC analyzed by \citet{Hld+08}. Different symbols refer
to different luminosity intervals, as explained in the legend. The microturbulent velocities $\microt$ have typical uncertainties of about
 $\pm 5\kms$. An uncertainty of 10\% or $\pm 10\kms$, whichever is the larger, 
should be considered for the rotational velocity measurements.}
  \label{vsinimicro}
\end{figure}

Photospheric microturbulence is routinely required, e.g., to derive consistent surface abundances
for one element from different photospheric absorption lines through stellar model atmospheres
\citep[among many others][]{1996A&A...315...95R,1998A&A...332..681H,2000A&A...358..639V}. 
Unfortunately, differences in physical assumptions or atomic physics can require
somewhat different microturbulent velocities for the same star in different studies. 
Here, we restrict our detailed comparison
to the data of \citet{Tdh+07} and \citet{Hld+08} from the ESO 
VLT-FLAMES Survey of Massive Stars \citep{Esl+05}, since 
it comprises the largest available uniformly analyzed data set.  
In Fig.\ref{vsinimicro}, we plot the microturbulent velocities derived for
the LMC early~B type stars analyzed by \citet{Hld+08} versus their 
projected rotational velocity. 
The error bar on the derived microturbulent velocities is usually quite big, 
$\pm 5\kms$, and is often comparable to the measured quantity itself. 
There seems to be no positive correlation between 
$\microt$ and the apparent projected rotational velocity
    $\vsini$. Though $\vsini$ is plotted and not 
    $\varv$ itself, the lack of a correlation in such a large 
    data set (justifying the assumption of random orientation of the
    sample) argues against rotation as an important effect in triggering 
    microturbulence in hot stars. To compare microturbulent velocities
    to properties of sub-photospheric convection we use only data
    obtained for slow rotators (i.e. $\vsini < 80\kms $)
    as microturbulent velocities are more difficult to measure for
    faster rotators.

\begin{figure}
  \resizebox{\hsize}{!}{\includegraphics{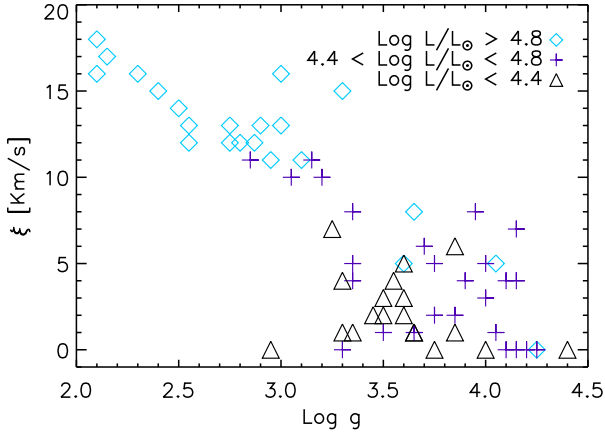}}
  \caption{Logarithm of surface gravity versus microturbulent velocity $\microt$ for the LMC early
B-type stars studied by \citet{Hld+08}; only stars with $\vsini < 80\kms$ are considered here. Different symbols refer to different luminosity intervals, as explained in the legend.
 The microturbulent velocities $\microt$ have typical uncertainties of about $\pm 5\kms$. For the surface gravity measurements 
an uncertainty of $\pm 0.1$ should be considered.} 
  \label{microlmc}
\end{figure}

In Fig.\ref{microlmc}, we show the microturbulent velocities for the LMC stars of \citet{Hld+08}
versus the stellar surface gravity. Trends of the  microturbulent velocity with $\log g$ have been
previously reported for hot stars \citep[e.g][]{Gl92,Hds+07}.
The figure shows that indeed, for $\log g < 3.2$, there is a clear trend. However,
the luminosity coding in Fig.\ref{microlmc} suggests that this trend may be largely produced by
the increase in convective velocity with increasing luminosity  (Sect.~\ref{results}).  
This figure displays a detection threshold of about $10\kms$ for 
the microturbulent velocities 
so in the following we restrict the comparison to $\microt \geq 10\kms$.

\begin{figure}
\resizebox{\hsize}{!}{\includegraphics[width=1.0\textwidth]{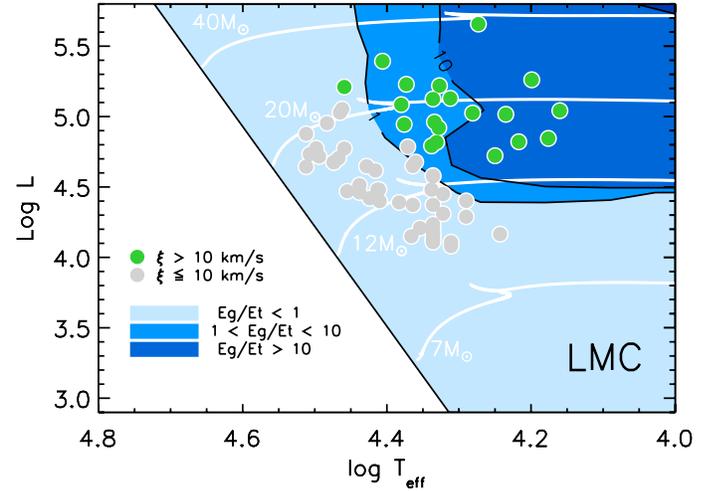}}
\caption{Values of the ratio $E_{{\mathrm g}}/E_{\mathrm s}$ of the kinetic energy in the form of gravity waves above the iron convection zone, 
to the kinetic energy of the surface velocity field, as a function of the location in the HR diagram (see color scale). This plot is based on
evolutionary models between 5$\mso$ and 100$\mso$ for LMC metallicity. We estimated the ratio $E_{{\mathrm g}}/E_{\mathrm s}$ as in Eq.~\ref{ratio}, 
using a value $\varv_s = 10\kms$ for the surface velocity amplitude. Over-plotted as filled circles are stars which have photospheric microturbulent 
velocities $\microt$ derived in a consistent way  by \citet{Hld+08}. Here, we use only data
for stars with an apparent rotational velocity of $\vsini < 80 \kms $. The uncertainty
in the determination of $\microt$ is typically $\pm 5\kms$, which justifies our choice of  $\varv_s = 10\kms$. 
Solid white lines are reference evolutionary tracks.
The full drawn black line corresponds to the zero age main sequence. }
  \label{grwplot}
\end{figure}

In order to compare these observations to our model predictions, we evaluated 
 the ratio 
of the kinetic energy in the form of gravity waves at the 
surface of the FeCZ to the kinetic 
energy of the surface velocity field,  $E_{\mathrm g}/E_{\rm s}$ 
(Eq.~\ref{ratio}), 
assuming $\varv_{\rm s} = 10\kms$, in the HR diagram.
Fig.~\ref{grwplot} shows two different iso-contours of this ratio; 
the stars of the LMC sample shown in Fig.~\ref{microlmc} are over plotted. 
Notably, all but one of the LMC stars of Fig.~\ref{grwplot} with $\microt > 10\kms$ 
are found in that part of the HR diagram where it is energetically
possible that the FeCZ-induced gravity waves trigger a significant
surface velocity field ($\varv_{\rm s} > 10\kms$). 
Thus, a physical connection of the FeCZ 
with the observed microturbulent
velocities appears energetically possible. Moreover, that the iso-contour line of
$E_{\mathrm g}/E_{\rm s} = 1$ in Fig.~\ref{grwplot} almost perfectly divides the
observed sample in stars with significant ($\microt > 10\kms$) and insignificant
($\microt < 10\kms$) microturbulence is a further 
indication of such a physical connection.

\begin{figure}
        \begin{minipage}[b]{0.5\textwidth}
                \centering
                \includegraphics[angle=0,width=1.0\textwidth]{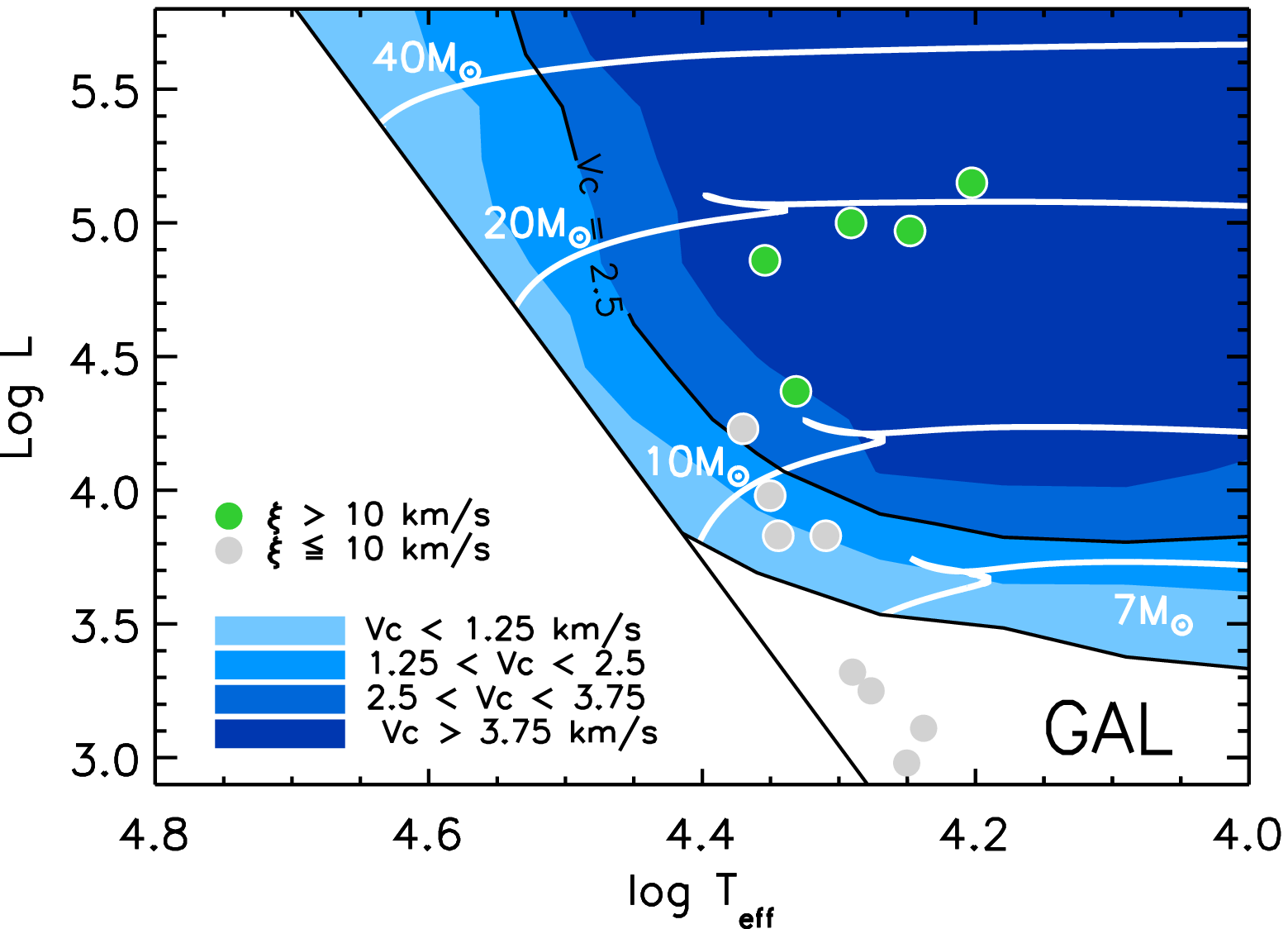}
                \includegraphics[angle=0,width=1.0\textwidth]{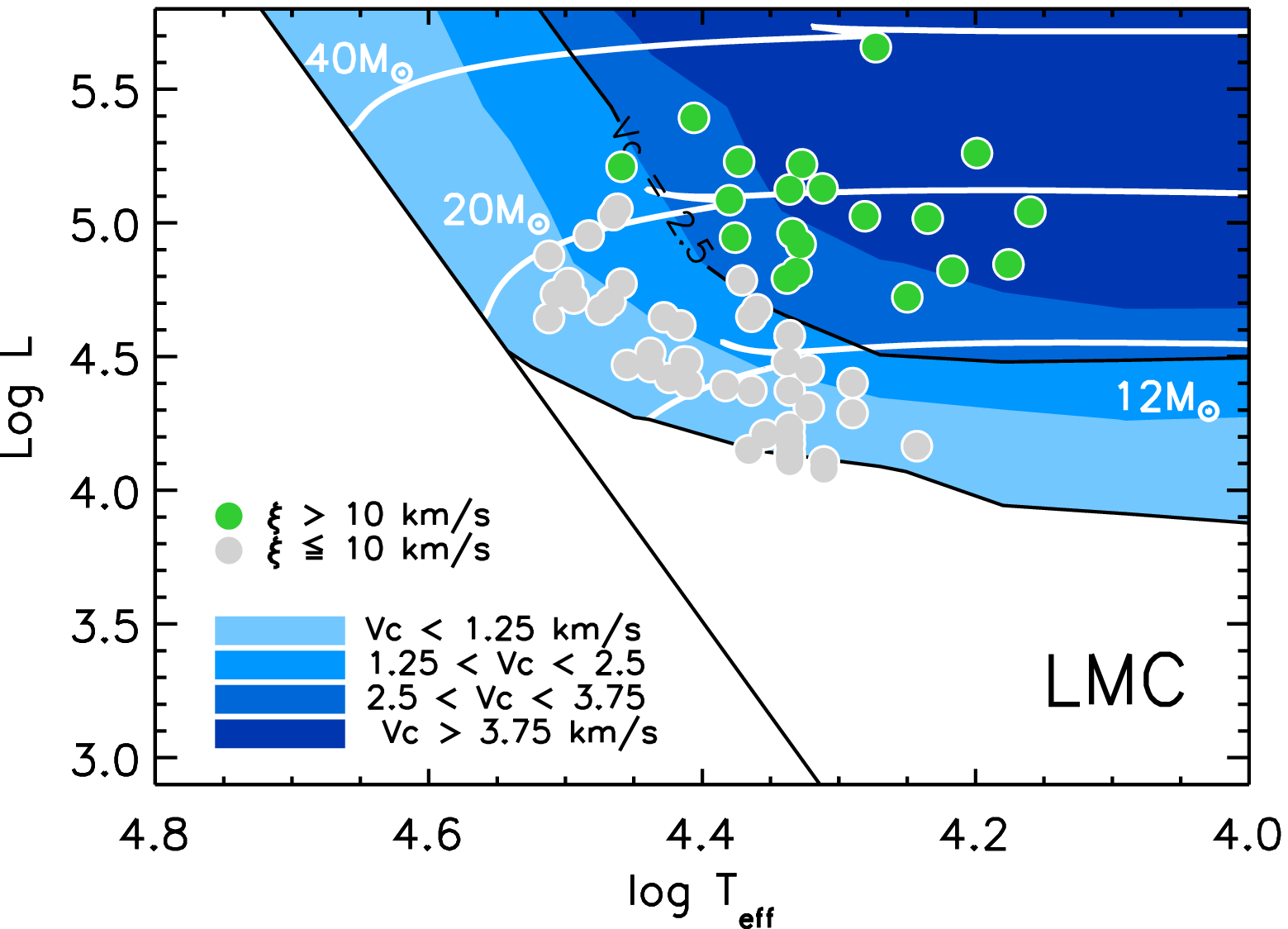}
        \end{minipage}
        \begin{minipage}[b]{0.5\textwidth}
                \centering
                \includegraphics[angle=0,width=1.0\textwidth]{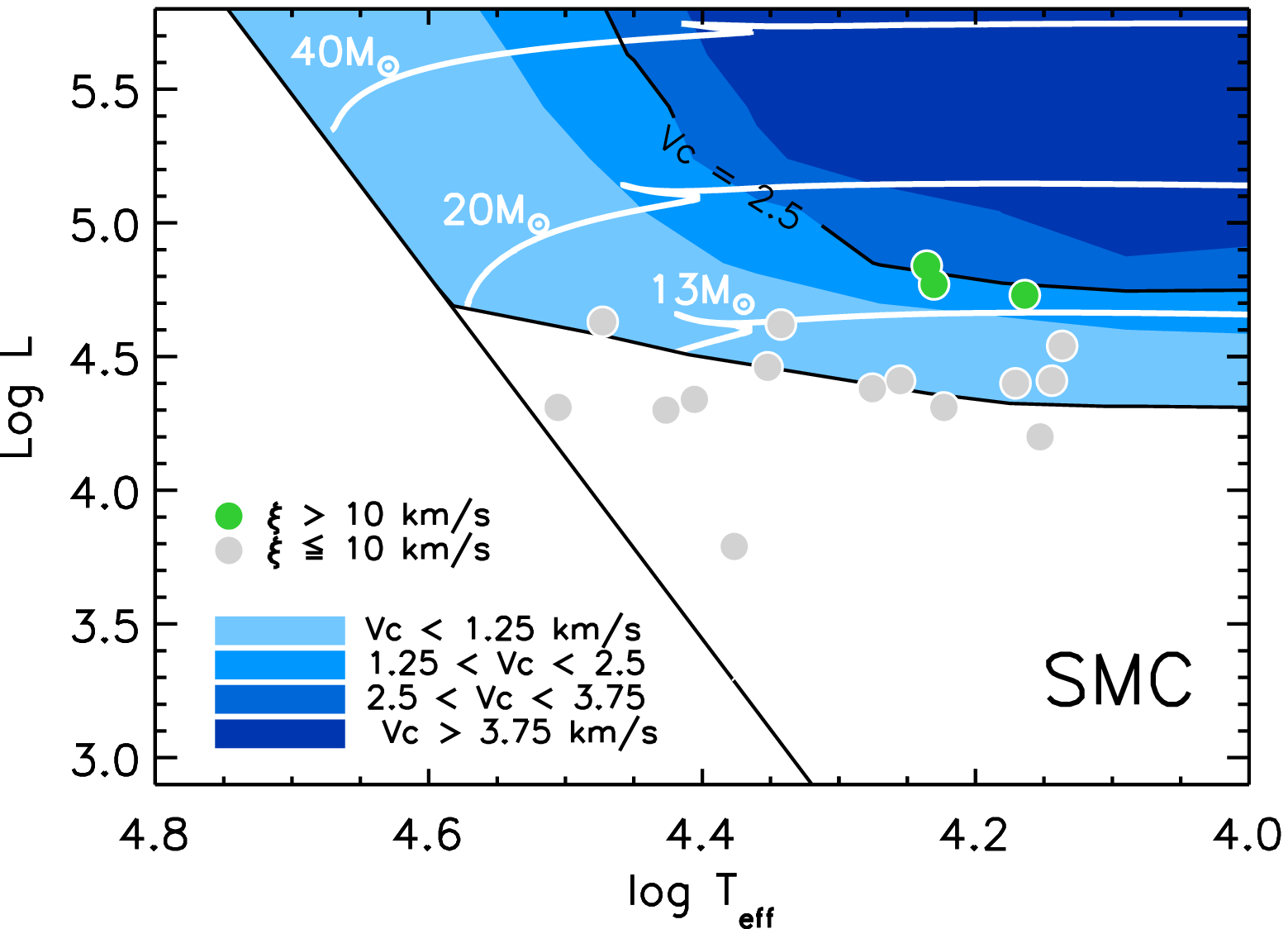}
        \end{minipage}
        \caption{Average convective velocity within 1.5 pressure scale heights of the
upper border of the iron convection zone in our models, as function of the location
in the HR diagram (see color scale), based on evolutionary models between 5$\mso$ and 100$\mso$
(white lines), for three metallicities corresponding to the Galaxy (top panel),  the LMC (middle),
 and the SMC (bottom). The full drawn black line corresponds
 to the zero age main sequence.
Over-plotted as filled circles are photospheric microturbulent velocities $\microt$ derived in a consistent way
for hot massive stars by  \citet{Tdh+07} and \citet{Hld+08}. Here, we use only data
for stars with an apparent rotational velocity of $\vsini < 80 \kms $. The uncertainty
in the determination of $\microt$ is typically $\pm 5\kms$.
}
\label{hrdvel}
\end{figure}

Figure~\ref{hrdvel} shows the iso-contours in the HR diagram of the  
average convective velocity from our models 
in the upper layers of the iron convective zone, $\vca$ (cf., Sect.~\ref{waves}), 
at the three considered metallicities.
We have over plotted the microturbulent velocities derived by 
\citet{Tdh+07} and \citet{Hld+08} as filled circles. 
Again, we distinguish between sample stars with significant ($\microt > 10\kms$; Group~A) and insignificant
($\microt < 10\kms$; Group~B) microturbulent velocities. 
Comparing the plot for the LMC in Fig.~\ref{hrdvel} with Fig.~\ref{grwplot} identifies 
$\vca \simeq 2.5\kms$ as a critical convection velocity to be able to trigger 
microturbulence. Interestingly, the contour
of  $\vca = 2.5\kms$ in our stellar models forms an almost perfect dividing line between Groups~A and~B
for all three considered metallicities.

In fact, Fig.~\ref{hrdvel} provides evidence for all three trends found in the average
convection velocity as function of stellar parameters (cf., Sect.~\ref{results}) to be present also in the
empirical data on microturbulent velocities.  The LMC data shows 
that in the luminosity range $4.5 < \llso < 5.5$ 
microturbulence is found only for $T_{\rm eff} \simle 25\,000\,$K.  
The data for all three metallicities clearly suggests a key role of the luminosity,
as the stars with $\microt > 10\kms$ are the most luminous ones in each sub sample.
And finally, the stars with high microturbulent velocities are all comfortably above
our theoretical contour line corresponding to $\vca = 2.5\kms$. 
As the latter trends toward the upper right corner of the 
HR diagram for lower metallicity, the metallicity dependence is also confirmed
by the empirical data.

\citet{Lrl04} studied microturbulence in a sample of 100 Galactic early B~stars.
Interestingly, they found significant microturbulent velocities (i.e., clearly above 5$\kms$) in the mass
range 7...11$\mso$ for stars with a relative age on the main sequence of $t/\tau_{\rm MS} > 0.8$,
and in the range 12...19$\mso$ for $t/\tau_{\rm MS} > 0.25$, but only insignificant
microturbulent velocities for younger or less massive stars. 
Again, these results appear to agree with Fig.~\ref{hrdvel} up to a remarkable 
quantitative  level. 

In summary, our comparison provides evidence for  a physical connection 
of microturbulence in hot star photospheres with the existence and strength of a sub-photospheric
FeCZ.  

If microturbulence has a physical origin and is not just a fudge factor,
    the pressure and energy terms associated with such a velocity field
    should be included in the calculations of atmospheric models of massive
    stars. \citet{1991ApJ...377L..33H} have investigated part of these effects by
    accounting for a constant microturbulent velocity in the pressure term
    only. They find that for stars with conspicuous $\microt$ values
    (of 25$\kms$) the inclusion of the pressure term leads to higher
    values of the surface gravity, which can reduce the mass discrepancy
    for O stars and O-type central stars of planetary nebula. 
    A similar approach was also studied by \citet{1998MNRAS.299.1146S}. 
    The impact on gravity discussed by \citet{1991ApJ...377L..33H} is likely an upper limit to the
    effect as, first, the $\microt$ values are in most cases less
    than 25$\kms$, and, second, a positive gradient in the 
    atmospheric $\microt(r)$ would decrease the pressure gradient due to 
    microturbulence but, to date, the radial
    stratification of the microturbulent velocity in the atmospheres of hot
    massive stars has not been studied in detail. From a theoretical 
    perspective, investigating $\microt(r)$ requires hydrodynamic simulations 
    of the stellar atmosphere, including the presence of sub-surface convection.
    
\begin{figure}
  \resizebox{\hsize}{!}{\includegraphics{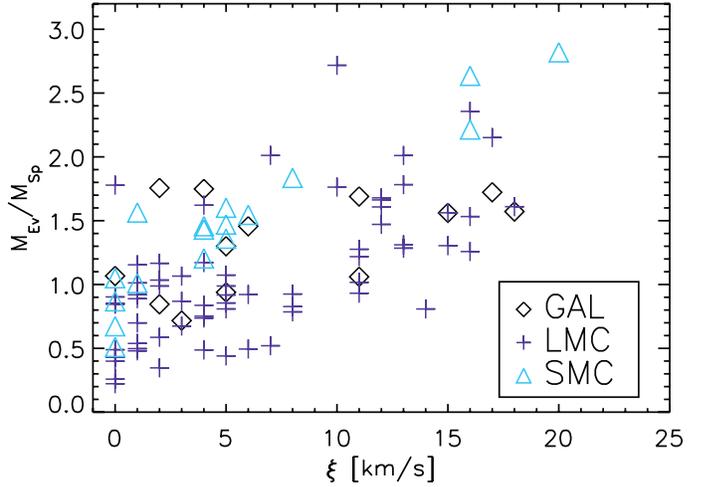}}
  \caption{Values of the mass discrepancy (evolutionary mass divided
  by spectroscopic mass) as function of microturbulent velocity in the
  sample of B stars analysed by  \citet{Tdh+07} and \citet{Hld+08}. 
   Here, we use only data for stars with an apparent rotational velocity of $\vsini < 80\kms $.}
  \label{discrepancy}
\end{figure}

    The mass discrepancy in massive stars is a well documented problem
    \citep[see for example][]{1992A&A...261..209H,2003A&A...398..455L,2005A&A...434..677T,2005ApJ...627..477M,2007A&A...465.1003M}.
    It is typically found that the difference between spectroscopic 
    mass and evolutionary mass is most pronounced in supergiants. In main
    sequence stars it may not be present at all, but see \citet{Hld+08}.
    Given that microturbulent velocities are highest in supergiants 
(see Fig.~\ref{microlmc}) an empirical correlation between 
    mass discrepancy and microturbulent velocity is to be expected and is 
    shown in Fig.~\ref{discrepancy} using data analysed by \citet{Tdh+07} and \citet{Hld+08}. 
    If indeed microturbulence is related to 
    subsurface convection and supergiants have intrinsically higher
    microturbulent velocities than dwarfs (see Section~\ref{results}) potentially
    part of the gradient in Fig.~\ref{discrepancy} may be explained by the effect discussed
    by \citet{1991ApJ...377L..33H}.

\subsection{Non-radial pulsation}\label{pulsation}

\subsubsection{Theoretical considerations}\label{stwaves}

In our discussion thus far we have considered 
only the propagation of running waves, 
it is possible that the stochastic convective 
motions can also excite standing waves,
i.e. high-order non-radial pulsations. For example, stochastic excitation 
is thought to be the cause
of the Solar oscillations \citep{1970ApJ...162..993U,1971ApL.....7..191L}. 
It may thus be possible that the FeCZ excites 
non-radial pulsations in hot early-type stars.

Several classes of OB~star models are found to be linearly unstable against non-radial
pulsations, among which are 
the $\beta~Cephei$ stars and the 
slowly pulsating B~stars \citep[e.g., ][]{2001A&A...366..166D,1999AcA....49..119P}.
The key ingredient required for the pulsational instability is the iron opacity
peak described in Sect.~\ref{method}. 
As convection is not required to produce the pulsations in these models,
it is not considered in detail as excitation mechanism \citep{2008arXiv0812.2149D}. 
It is conceivable that the convective excitation could modify the predicted pulsation
spectrum and/or extend the instability region of certain linear instabilities. 
The convective kinetic energy flow into waves 
could be predominantly directed into those modes for 
which instability is predicted in the models. 
In certain parts of the HR diagram, one may thus 
suspect an intricate connection between
the occurrence of a sub-photospheric iron convection 
zone and the properties of non-radial pulsations.

Non-radial pulsations have also been considered as the origin of various observed
small scale \citep[e.g., line profile
 variability, ][]{1996ApJS..103..475F,1997A&A...327..699F} and large scale phenomena 
\citep[e.g., so called discrete absorption
  components, ][]{1988MNRAS.233..123P,1995ApJ...452L..53M,1997A&A...327..281K,2002A&A...388..587P}
 at the surface or in the wind of massive OB stars.
Non-radial g-mode pulsations were also recently proposed as the origin
of observable macroturbulence in massive B~type stars \citep{2008arXiv0812.2641A}.
In Fig.~\ref{pulsationplot} we compare the regions where strange mode, g-mode, 
and p-mode pulsations are predicted to occur in the HR diagram 
with the region where our models predict a strong FeCZ. 
Pulsations appear to be almost ubiquitous when all types of variables are accounted
for. The strange mode pulsators are predicted to cover the HR diagram at high
luminosity, where we plotted only the predictions for the radial strange modes of \citet{1993MNRAS.264...50K}; 
high-order non-radial strange modes seem to be omnipresent
as well for stars above 40$\mso$ or so \citep{1996MNRAS.282.1470G}.
Non-radial g-mode pulsators  are predicted by \citet{2006ApJ...650.1111S}
in the B~supergiant region. And radial and low order non-radial p-modes are predicted for the
$\beta\,$Cephei regime by \citet{2001MNRAS.327..881D} and by \citet{1999AcA....49..119P} and
\citet{2006ApJ...650.1111S} for a considerably larger region in the HR diagram.
At lower metallicity, many of the predicted areas in the HR diagram are 
smaller \citep[cf.,][]{1993MNRAS.264...50K, 2001MNRAS.327..881D} but 
the general picture is still incomplete.

\subsubsection{Observations}

\begin{figure}
  \resizebox{\hsize}{!}{\includegraphics{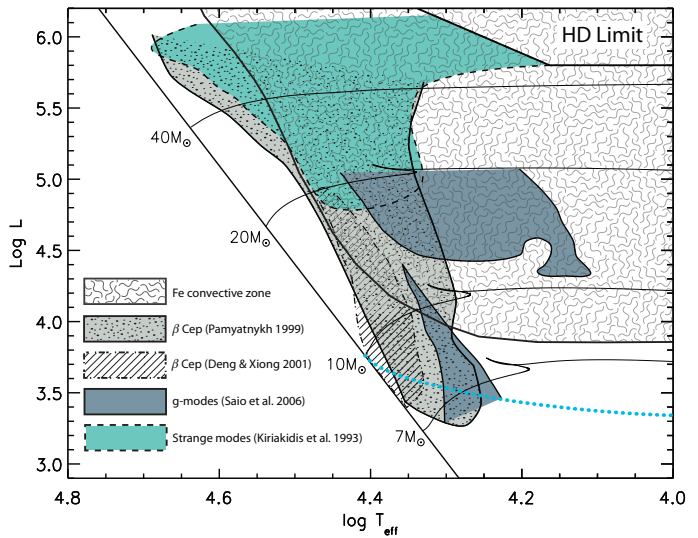}}
  \caption{The plot shows regions of the HR diagram where pulsational instabilities are predicted, compared to our calculations for the occurrence
   of iron convection. The cloudy region  marks the presence of iron sub-surface convection with $\vca \geq 2.5\kms$, while the dotted, blue line 
   divides regions of the HR diagram where iron convection is present (above) from regions where it is absent (below). Different modes of instabilities are shown
   with different colors and different contour line styles, as explained in the legend. Evolutionary tracks between 7$\mso$ and 40$\mso$ are 
   plotted as a reference. The straight, full drawn black line
   corresponds  to the zero age main sequence. The
   Humphrey-Davidson limit is also plotted for reference (top-right corner).}
  \label{pulsationplot}
\end{figure}

Observationally, the classical $\beta\,$Cephei stars are concentrated in the
region predicted by Deng \& Xiong \citep{2005ApJS..158..193S}, while the
B~supergiant non-radial g-mode pulsators overlap with the prediction of \citet{2006ApJ...650.1111S}
but extend to an effective temperature of $\sim 10\, 000\,$K \citep{2007A&A...463.1093L}.
Pulsations are also found for the most luminous stars \citep[e.g., the $\alpha$ Cygni-variables;][]{1998A&AS..128..117V}, 
but there is now no clear evidence for strange mode pulsators.  
Comparing the prediction for the FeCZ 
with that for pulsational instability
(Fig.\ref{pulsationplot}) shows two things. Firstly, the FeCZ-region is much larger than any region for
a particular pulsational instability. Thus, distinguishing whether a certain observational
feature is caused by a particular pulsational instability by the FeCZ might, in principle,
be possible, since the area in the HR diagram where the latter occurs but
the pulsational instability does not is relatively large. 
Secondly, some regions exist where (so far) no pulsations are predicted but the FeCZ in our
models is strong, or where, vice versa, pulsations are predicted but the FeCZ is weak or absent.

Comparing Fig.~\ref{pulsationplot} with Fig.~\ref{hrdvel}, where we 
show the observations of microturbulence and 
the FeCZ predictions, it is unlikely that microturbulence
is associated with a particular pulsational instability. Strong microturbulence
is observed at too low a luminosity to be attributable to strange mode
pulsations alone, while p-mode pulsators are found where microturbulence seems not to occur.
Concerning the g-mode pulsators the situation is less clear. Fig.~\ref{pulsationplot} shows that, at
solar metallicity, g-mode pulsations for post-main sequence stars are expected
only in a rather narrow luminosity interval. Unfortunately, the five Galactic stars
shown in Fig.~5 for which strong microturbulence is derived are all inside this
luminosity range, so they cannot distinguish between a pulsational or FeCZ origin
of microturbulence. However, looking at the LMC data, stars above the
g-mode luminosity upper limit with microturbulence are found; whether or not corresponding
stellar models are g-mode unstable is currently not known. 
A connection of microturbulence with non-radial pulsations is thus not impossible,
but it is also not very likely. 

Comparing Fig.~\ref{pulsationplot} with the discrete absorption components (DACs) found in 200~Galactic O~stars above $\sim 20\mso$
by \citet{1989ApJS...69..527H} all the way to the zero age main sequence,
seems to argue against non-radial pulsations as the origin of the DACs phenomenon
(see also Sect.~\ref{magnetic}). 

\subsection{Wind clumping}\label{clumping}

\subsubsection{Theoretical considerations}\label{tclump}
Observational evidence exists for stellar wind inhomogeneities on small
and on large scales. While the latter will be discussed in Sect.~\ref{magnetic},
here we consider only small scale wind structure, or wind clumping.
In Sect.~\ref{micro}, we discussed that waves produced by the FeCZ could lead to velocity fluctuations at the stellar surface.
In order to induce wind clumping, those waves should induce density fluctuations
at the stellar surface. Through the occurrence of porosity or shifts in the ionisation
balance of the gas the mass-loss rate may be affected. 
For this to happen, the amplitude of the velocity 
fluctuations at the surface should be on the same order of the sound speed. 
Alternatively, the velocity fluctuations
might directly affect the local mass-loss rate through the Doppler effect,
if the amplitude of the velocity fluctuations is on the same order
as the speed of the wind flow, which, at the base of the wind, is approximately 
the sound speed. As the sound speed at the surface in our
massive main sequence models is on the order of a few times $10\kms$,
we consider here those stellar models potentially capable to produce wind
clumping for which the convective velocities
in the upper part of the FeCZ $\vca \geq 2.5\kms$,
as this allows energetically to have surface velocity amplitudes
above $\sim 10\kms$ (cf. Sect.~\ref{micro}).   

Assuming the horizontal extent of the clumps to be comparable to the
sub-photospheric pressure scale height $H_{\rm p}$, we may estimate
the number of convective cells by dividing the stellar surface area
by the surface area of a convective cell finding that it scales
with ($R/H_{\rm P})^2$. For our main sequence O star models in the mass
 range 20-60 $\mso$, we find pressure scale heights in the range 0.04-0.24
$\rso$, corresponding to a total number of clumps in the range 6 $\times 10^3-
6 \times 10^4$. In principle, this might be testable through
linear polarization variability measurements, which can probe wind asphericity 
at the very base of the wind \citep{2007A&A...469.1045D}. 

\subsubsection{Observations}

Evidence has been accumulating that the winds of massive stars 
may be subject to small scale clumping. So far this is best
documented for Wolf-Rayet (WR) stars, where line variability on time 
scales of minutes to hours is thought to constitute direct evidence of 
outflows that are clumped already in the acceleration zone near the 
base of the wind \citep{1999ApJ...514..909L}. This clumping may be
part of the explanation for the wealth of intricate detail seen in
nebulae around WR stars \citep{1998ApJ...506L.127G}. Recently, 
\cite{2008arXiv0805.1864L} reported spectroscopic variability in 
the Of supergiants $\zeta$\,Pup \citep[see also][]{1998ApJ...494..799E} 
and HD\,93129A. The amplitude of the variation (at the 1-3\% level) is 
similar as in WR stars supporting the notion that clumping is not 
restricted to WR stars.

Indeed, evidence that O star winds are clumped is given by, among others,
\cite{2006A&A...454..625P}. These authors investigate the clumping
behavior of the inner wind (inside about two stellar radii) relative
to the clumping in the outer wind (beyond tens of stellar radii)
of a large sample of supergiant and giant stars.
They find that in stars that have strong winds, the inner 
wind is more strongly clumped than the outer wind, whereas those 
having weak winds have similar clumping properties in the inner and 
outer regions. Their analysis only allows for such a relative statement.
In principle, for weak winds the outer part could be homogeneous.
If so, weak winds are not clumped.
In any case, strong winds - identified as such if H$\alpha$ is seen in 
emission - are clumped near the base of the wind.  
A measure of the degree of clumping is the clumping factor $f_{\rm cl} =
     \langle \, \rho^{2}\rangle/ \langle \, \rho\rangle^{2} \geq 1$ where angle brackets denote (temporal) average
     values \citep[e.g.][]{2006A&A...454..625P}.
     
Apparently, this type of radial behavior is not consistent with
hydrodynamical predictions of the intrinsic, {\em self-excited}
line-driven instability \citep{2002A&A...381.1015R,
2005A&A...429..323R}. Such models predict a lower clumping in the
inner wind than the outer wind. Moreover, if there was any dependence
on wind density predicted at all, optically thin winds should be more strongly
clumped than optically thick winds \citep{1999ApJ...510..355O,2006A&A...454..625P}. 
Therefore, the findings on the radial clumping behavior in O stars
may point to an additional excitation mechanism of wind structure. 

\begin{figure}
    \resizebox{\hsize}{!}{\includegraphics{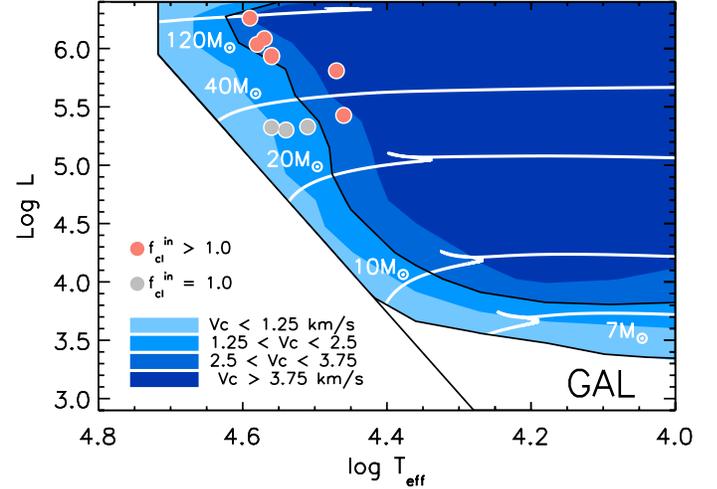}}
  \caption{Average convective velocity within 1.5 pressure scale heights of the
upper border of the iron convection zone in our models, as function of the location
in the HR diagram (see color scale), based on evolutionary models between 5$\mso$ and 120$\mso$
(white lines) at solar metallicity. The full drawn black line corresponds  to the zero age main sequence.
Over-plotted as filled circles are observations of the clumping factor $f_{cl}^{in}$ 
(see text for definition) in the winds of O stars, according to 
\citet{2006A&A...454..625P}. The data shown here corresponds to objects 
with well-constrained clumping parameters.
Note the different luminosity range with respect to Fig.~\ref{hrdvel}.
 }
  \label{clumpvel}
\end{figure}

Fig.~\ref{clumpvel} shows that
the O stars investigated by \cite{2006A&A...454..625P} populate the
regime in the HR diagram
in which our models predict the average convective velocity in the
top part of the FeCZ to change from a few to 
over $2.5\kms$, indicating that surface velocity fluctuations
on the order of the local sound speed are possible (cf. Sect.~\ref{tclump}). 
Though the part of the HR diagram that is covered by the
sample is limited ($4.46 \la \log\,T_{\rm eff} \la 4.66$; $5.29 \la
\log\,L/L_{\odot} \la 6.26$), the trend is such that stars with
relatively strong clumping in the inner winds are in a regime where
$\vca$ is higher. A correlation between clumping 
at the base of the wind and $\vca$,
i.e., between wind clumping and the properties of the FeCZ,
appears therefore possible.
To further test the idea that the FeCZ
 produces wind clumping at the wind base for sufficiently luminous and cool stars
it would be desirable to derive the radial clumping profiles for cooler (i.e. B-type) stars.
If correct, such stars, both the ones with weak and
strong winds, should have relatively strong clumping at the base
of the wind.

To derive the spatial scale of the wind clumps from linear polarimetry has not yet been
possible for main sequence OB stars. A limitation 
is that this technique requires very high signal-to-noise observations 
\citep[see discussion in][]{2002MNRAS.337..341H}. Luminous Blue Variables (LBVs) however provide 
a more appropriate category of objects to test wind clump sizes, because of a 
combination of higher mass-loss rates, and lower wind velocities than for O stars 
\citep{2005A&A...439.1107D}. Indeed,
\citet{2007A&A...469.1045D} show that in order to produce the observed 
polarization variability of P~Cygni, the wind 
should consist of about $\sim$ 1000 clumps per wind flow-time ($\flt \equiv \rst/\vin$). 
To see whether this observational result is compatible with sub-surface convection
causing wind clumping,  
we considered the sub-surface convective regions of a massive star model with global 
properties similar to those of P Cygni (initial mass 60 $\mso$, $\log(L/L_{\odot}$ = 5.9, and 
$\teff$ = 18\,000 K ).  As a result of the lower gravity, 
the pressure scale height in the FeCZ in this model
is about 4$\rso$, which is much bigger than in our O star models. 
Consequently, the same estimate for the number of clumps as done for the main sequence
models in Sect.~\ref{tclump} yields about 500 clumps per wind flow time, a
number which is quite comparable 
to that derived for P~Cygni observationally (about $10^3$ clumps per
wind flow time). \\  

Finally,
\citet{1996ApJS..103..475F} have conducted a spectroscopic survey of O stars and observed intrinsic absorption
line profile variability (LPVs) for about 77\% of their sample. They report an increase of incidence and
amplitude of variability with increasing stellar radius and luminosity,
as well as no statistically significant
line profile variability for dwarfs earlier than O7.
While Fullerton et al. attempt to relate their findings to
the predictions of strange-mode pulsation in O~stars by
\citet{1993MNRAS.264...50K}, a comparison of their results
(see their Fig.~13)
with the occurrence of sub-surface convection as depicted
in Fig.~\ref{hrdvel} indicates the possibility
of a physical connection between line profile variability
and sub-surface convection in O~stars.

\subsection{Magnetic fields}\label{magnetic}

\subsubsection{Fields from iron convection zones}

In solar-type stars, surface convection zones modified by the stellar rotation
are thought of being capable of producing a
magnetic field through the so called  $\alpha\Omega$-dynamo \citep{1975ApJ...198..205P,1980Natur.287..616S,1992A&A...265..106S}.
The FeCZ in our massive main sequence stellar models has a spatial extent similar to the
Solar convection zone, although its mass is much smaller, and OB~stars are rapid rotators, so it is possible that a
dynamo may also work in the envelopes of~OB stars. 
If so, the magnetic field may be able to
penetrate the radiatively stable layer above the FeCZ, and 
dynamically significant field strengths might be achievable. 
To this end, we follow the model by \citet{2003ApJ...586..480M}
 for the rise of buoyant magnetic flux tubes
generated at the edge of the convective core of massive stars through their radiative
envelope and apply this model to the FeCZ and the overlying radiative 
layer. The magnetic field strength $B_0$ in the iron convection
zone is estimated assuming equipartition of 
kinetic energy density and magnetic energy 
density inside the convective layers:
\begin{equation}
B_0 \simeq  2\varv_c\sqrt{\pi \rho}, 
\end{equation}
which, for our 60$\mso$ star at $Z$=0.02, reaches about $B_0 \simeq  2700\,$G 
inside the iron convective zone.  The surface field $B_{\rm s}$ is then
obtained by multiplying this number with the ratio of the surface density $\rho_{\rm s}$ and the
density in the FeCZ $\rho_0$, i.e.  
$B_{\rm s} \simeq B_0 \, \rho_{\rm s} / \rho_0 \simeq 60\,{\rm G}$.
Similarly, for the 20$\mso$ model at $Z$=0.02 we obtain $B_0 \simeq 1400\,$G 
and $B_{\rm s}\simeq 10\,{\rm G}.$ 
Although at the surface, the magnetic pressure in the flux tubes is only on the order
of a few percent of the total pressure, it is on the same order as the gas pressure
and could thus lead to considerable horizontal density differences. 
Compared to the situation envisioned by \citet{2003ApJ...586..480M},
who found that the rise time of the flux tubes from the edge of the convective
core to the stellar surface can be comparable to the 
main sequence life time \citep[but see also][]{2004MNRAS.348..702M},
the rise time of the flux tubes from the FeCZ to the surface 
is much shorter. And while the initial magnetic field strength at the edge of
the convective core can be considerably higher than our values of $B_0$,
the surface fields obtainable from the sub-surface convection zones are higher,
due to the much lower density contrast between convection zone and surface in this case.

As a consequence, even 
though we are far from a detailed picture, it seems conceivable that
the FeCZs in massive main sequence stars produce localized magnetic
fields at their surface. 
The interaction of the stellar wind with the localized surface magnetic
fields could enhance the rate at which the wind induces a loss of stellar angular momentum.
Furthermore, co-rotating density patterns in the outflowing wind could be produced by these local magnetic spots.

Rotation may play an important role in the dynamo process, possibly resulting in the appearance of stronger fields 
at the surface for faster rotating stars. To estimate this effect, a dynamo model accounting for the differential rotation needs to be 
implemented in the stellar evolution calculations. This will be the discussed in a subsequent paper.

\subsubsection{Observations}

Surface magnetic fields have been linked to several observed phenomena 
in OB stars, e.g. discrete absorption components 
(DACs) in UV resonance lines
\citep[e.g.,][]{1988MNRAS.233..123P,1995ApJ...452L..53M,1997A&A...327..281K,2002A&A...388..587P}, 
which are thought to diagnose large scale coherent wind
anisotropies \citep{1996ApJ...462..469C,2008ApJ...678..408L},
or the less coherent line profile variability mentioned above \citep{1996ApJS..103..475F,1997A&A...327..699F}.
Also non-thermal X-ray emission of OB main sequence stars has been proposed to relate
to surface magnetic fields \citep[e.g.,][]{1997ApJ...485L..29B,2002ApJ...576..413U}.

A connection of the FeCZ in massive stars
with the phenomena mentioned above has not yet been considered. 
However, such a connection becomes testable through our results. 
While in our comparison to observed microturbulence presented above
we discussed when sub-surface convection may lead to detectable surface velocity
fluctuations, the presence of surface magnetic fields may simply depend on
whether an FeCZ is present in the star or not. 
Looking at Fig.~\ref{hrdvel}, we see that in our models the FeCZ 
is absent for luminosities below about $10^{3.2}\lso$, $10^{3.9}\lso$, and $10^{4.2}\lso$ 
for the Galaxy, LMC and SMC, respectively. If DACs or line profile variability
were produced by magnetic flux tubes generated in the FeCZ,
those phenomena would not be expected for OB stars below those luminosities.
\citet{1989ApJS...69..527H} find DACs in nearly  all O stars (97\%)
in a large Galactic sample, with $\llso > 4.5$ and with effective temperatures
as high as the zero-age main sequence values of stars above $\sim 20\mso$.
Since those stars are well above the luminosity threshold for the occurrence of the
iron convection at Galactic metallicity, these observations do not
exclude DACs being due to FeCZ induced B-fields.
Also, all eleven early B~supergiants with DACs in the sample
of \citet{2002A&A...388..587P}
are predicted to have strong FeCZ by our results.  
Notably, between about $20\mso$ and $40\mso$, stars
close to the zero-age main sequence are not predicted to be pulsationally
unstable (cf. Fig.~\ref{pulsationplot}), which may be in conflict with 
pulsations as the origin for DACs.

\subsubsection{Other types of fields}
It may be interesting to briefly compare the expectation
from surface magnetic fields produced via the FeCZ to that for 
fields produced by other means.  Surface fields produced by convective cores
 \citep{1978A&A....68...57S,2001ApJ...559.1094C,2003ApJ...586..480M} have been proposed to relate to the same phenomena as those
mentioned above,  even if for massive stars the buoyant rise of magnetic fields from the convective core seems to be unlikely  \citep{2004MNRAS.348..702M}.
In contrast to the sub-surface FeCZ, convective cores
are prevalent in all stars above about 1.2$\mso$. It has been found
that the longer lifetime of stars of lower
mass may favor the drift of fields produced in the core to the surface \citep{1978A&A....68...57S,2003ApJ...586..480M}. 
Therefore, the expected trend is opposite to that found for fields produced
by the FeCZ, where surface fields may occur only for stars
above a critical mass (or luminosity), and stronger fields are found for
more massive stars.

On the other hand, in contrast to fields from the FeCZ,
magnetic flux tubes produced in the core may carry CNO-processed 
material to the surface.
This might thus constitute a 
mechanism to explaining nitrogen enrichment in slowly rotating
early~B stars \citep{Mba+06,Mgb08,2008ApJ...676L..29H}.    
Strong fossil magnetic fields are thought to persist 
in only a fraction of massive stars 
\citep{2005MNRAS.356..615F,2004Natur.431..819B},
 and may lead to, among other phenomena,
highly anomalous surface chemical compositions, wind confinement, and variable 
X-ray emission \citep[e.g.,][]{2006A&A...451..195W,2005ApJ...630L..81T}. 
Those strong features can clearly not be produced by fields originating 
from the FeCZs.

Finally, magnetic fields produced in differentially rotating massive stars by the
Spruit-Taylor dynamo \citep{Spr02} 
may transport angular momentum and chemical species \citep[cf.,][]{hws05}.
These fields are predominantly of toroidal geometry and would quickly decay near the stellar surface,
and are thus not thought to lead to observable fields at the stellar surface (but see also \citet{2005MNRAS.356.1139M}).

\section{Concluding remarks}

Hot luminous stars show a variety of phenomena  at their photosphere and in their winds
which still lack a clear physical interpretation at this time. Among these phenomena are
photospheric turbulence, spectral line variability (DACs and LPVs; see Sect.~\ref{comparison}), wind
clumping, variable or constant non-thermal X-ray and radio emission, chemical composition
anomalies, and intrinsic slow rotation. In the previous section, we argued that the iron 
convection zone could be responsible for various of these phenomena. 

We proposed in Sect.~\ref{micro} that a physical connection may exist between
microturbulence in hot star atmospheres and a sub-surface FeCZ. 
The strength of the FeCZ is predicted to increase 
with increasing metallicity $Z$,  
decreasing effective temperature $T$ and increasing 
luminosity $L$ (Sect.~\ref{results}), and all three predicted trends  
are reflected in the observational data.
This suggests that microturbulence corresponds to a physical motion of
the gas in hot star atmospheres. This motion may then be  connected to
wind clumping (Sect.~\ref{clumping}), since the empirical 
microturbulent velocities are comparable to the local sound speed at the stellar surface. 
In order to verify such a picture, multi-dimensional calculations of the
FeCZ and the radiative layers above, including the stellar atmosphere,
are required --- similar to the recent generation of atmosphere models for cool stars 
\citep[e.g.,][]{1999A&A...346L..17A,2004A&A...414.1121W}.

In Sect.~\ref{magnetic}, we proposed that the FeCZ in hot stars might also produce
localized surface magnetic fields, in Galactic stars for luminosities above
$\sim 10^{3.2}\lso$. This could explain the occurrence of DACs (discrete absorption
components in UV absorption lines), also in very hot main sequence stars for which pulsational
instabilities are not predicted. 
We further argued that there may be  regions of the upper HR diagram for which the presence of
the FeCZ influences, or even excites, non-radial
stellar pulsations (Sect.~\ref{pulsation}). 

The FeCZ could also turn out to directly affect the evolution of hot massive stars.
If it induces wind clumping, it may alter the stellar wind
mass-loss rate. 
Such a change would also influence the angular
momentum loss. In addition  magnetic fields produced by the iron 
convection zone could lead to an enhanced rate of angular momentum loss.
These effects become weaker for lower metallicity,
where the FeCZ is less prominent or absent (see Sect.~\ref{results}). 

Finally, we note that the consequences of the FeCZ might be strongest
in Wolf-Rayet stars. These stars are so hot that the iron opacity peak, 
and therefore the FeCZ, can be directly at the stellar surface,
or --- to be more precise --- at the sonic point of the wind flow \citep{hl96}. 
This may relate to the very strong clumping found observationally
in Wolf-Rayet winds \citep{1999ApJ...514..909L,2006ApJ...639L..75M}, and may be required for an understanding of the
very high mass-loss rates of Wolf-Rayet stars \citep{1995ApJ...448..858E,ki92,hl96}.

\begin{acknowledgements} 
The authors would like to thank Myron Smith, Alex Fullerton, Derk Massa and the members
of the VLT-FLAMES consortium. MC wants to thank the STScI for kind hospitality and the Leids Kerkhoven-Bosscha Fonds 
for financial support. S.-C. Y. is supported by  the DOE SciDAC Program (DOE DE-FC02-06ER41438).
\end{acknowledgements}

\bibliographystyle{aa} 
\bibliography{references}

\Online
\begin{appendix}

\onecolumn
\section{Models}
\begin{center}
\begin{longtable}{cccccccccc}

\caption[Table]{Outermost 4 $\rso$ of a 60 $\mso$ model at solar metallicity ($Z$=0.02). The table shows  physical variables at t=\tsixty years. Columns 
contain the progressive grid point of the model, the status (R=radiative, C=convective), the optical depth $\tau$, the opacity $\varkappa$,
the density $\rho$, the radius R, the value $M_{*}-M_r$ (where $M_*$ is the total stellar mass and $M_r$ is the mass coordinate), the temperature T,
the convective velocity $v_c$ and the local sound speed $c_s$. All the values are in cgs units if not otherwise specified. 
} \label{table60} \\

\hline\hline
\\[-2ex]
  
 Grid point &  STAT &  $\tau$ &  $\varkappa$ &  $\rho$ &    $R$ [$\rso$] &      $M_{*}-M_r$ [g] &  T[K] &       $v_c$ [km~s$^{-1}$] &  $c_s$ [km~s$^{-1}$] \\
\\[-2ex]
\hline
\endfirsthead

\\[-2ex]
\hline\hline
 Grid point &  STAT &  $\tau$ &  $\varkappa$ &  $\rho$&    $R$ [$\rso$] &      $M_{*}-M_r$ [g] &  T[K] &       $v_c$ [km~s$^{-1}$] &  $c_s$ [km~s$^{-1}$] \\
\\[-2ex]
\hline

\endhead

  \multicolumn{3}{l}{{Continued on Next Page\ldots}} \\
\endfoot

\\[-2ex]
\hline 
\endlastfoot

\hline

1062 & R & 1.82E+03 &    0.88736 &  1.2633E-08 &    18.0541 & 2.14E-05 & 2.240E+05 &       0.00 &     267.57 \\
 1063 & R & 1.76E+03 &    0.89553 &  1.1674E-08 &    18.1381 & 2.06E-05 & 2.214E+05 &       0.00 &     271.52 \\
 1064 & R & 1.70E+03 &    0.90534 &  1.0813E-08 &    18.2281 & 1.99E-05 & 2.187E+05 &       0.00 &     275.00 \\
 1065 & R & 1.64E+03 &    0.91747 &  1.0071E-08 &    18.3243 & 1.92E-05 & 2.159E+05 &       0.00 &     277.52 \\
 1066 & R & 1.57E+03 &    0.92978 &  9.5763E-09 &    18.4264 & 1.84E-05 & 2.136E+05 &       0.00 &     278.41 \\
 1067 & R & 1.53E+03 &    0.94138 &  9.2743E-09 &    18.4932 & 1.79E-05 & 2.118E+05 &       0.00 &     278.04 \\
 1068 & C & 1.49E+03 &    0.95487 &  9.0395E-09 &    18.5618 & 1.75E-05 & 2.099E+05 &      15.64 &     276.68 \\
 1069 & C & 1.45E+03 &    0.96881 &  8.8184E-09 &    18.6315 & 1.70E-05 & 2.080E+05 &      27.92 &     275.11 \\
 1070 & C & 1.41E+03 &    0.98203 &  8.6221E-09 &    18.7025 & 1.66E-05 & 2.062E+05 &      34.25 &     273.50 \\
 1071 & C & 1.37E+03 &    0.99441 &  8.4477E-09 &    18.7644 & 1.62E-05 & 2.046E+05 &      38.62 &     271.88 \\
 1072 & C & 1.34E+03 &    1.01310 &  8.1959E-09 &    18.8272 & 1.58E-05 & 2.020E+05 &      43.63 &     269.23 \\
 1073 & C & 1.26E+03 &    1.03152 &  7.9584E-09 &    18.9554 & 1.50E-05 & 1.994E+05 &      47.93 &     266.34 \\
 1074 & C & 1.23E+03 &    1.04388 &  7.8068E-09 &    19.0207 & 1.46E-05 & 1.977E+05 &      50.34 &     264.25 \\
 1075 & C & 1.19E+03 &    1.05647 &  7.6607E-09 &    19.0868 & 1.42E-05 & 1.959E+05 &      52.58 &     262.03 \\
 1076 & C & 1.15E+03 &    1.06922 &  7.5199E-09 &    19.1537 & 1.38E-05 & 1.941E+05 &      54.67 &     259.68 \\
 1077 & C & 1.11E+03 &    1.08107 &  7.3950E-09 &    19.2214 & 1.34E-05 & 1.924E+05 &      56.42 &     257.39 \\
 1078 & C & 1.08E+03 &    1.09196 &  7.2848E-09 &    19.2792 & 1.30E-05 & 1.908E+05 &      57.92 &     255.19 \\
 1079 & C & 1.05E+03 &    1.10822 &  7.1261E-09 &    19.3375 & 1.27E-05 & 1.884E+05 &      60.52 &     251.67 \\
 1080 & C & 9.84E+02 &    1.12405 &  6.9774E-09 &    19.4556 & 1.20E-05 & 1.860E+05 &      62.47 &     247.93 \\
 1081 & C & 9.52E+02 &    1.13434 &  6.8840E-09 &    19.5154 & 1.17E-05 & 1.843E+05 &      63.23 &     245.26 \\
 1082 & C & 9.19E+02 &    1.14425 &  6.7952E-09 &    19.5756 & 1.13E-05 & 1.826E+05 &      64.34 &     242.46 \\
 1083 & C & 8.87E+02 &    1.15367 &  6.7112E-09 &    19.6362 & 1.10E-05 & 1.809E+05 &      65.35 &     239.53 \\
 1084 & C & 8.54E+02 &    1.16168 &  6.6393E-09 &    19.6972 & 1.07E-05 & 1.793E+05 &      66.06 &     236.74 \\
 1085 & C & 8.27E+02 &    1.16833 &  6.5784E-09 &    19.7473 & 1.04E-05 & 1.779E+05 &      66.61 &     234.13 \\
 1086 & C & 8.00E+02 &    1.17736 &  6.4933E-09 &    19.7975 & 1.01E-05 & 1.757E+05 &      68.07 &     230.00 \\
 1087 & C & 7.47E+02 &    1.18517 &  6.4174E-09 &    19.8986 & 9.56E-06 & 1.734E+05 &      68.74 &     225.65 \\
 1088 & C & 7.20E+02 &    1.18966 &  6.3727E-09 &    19.9493 & 9.28E-06 & 1.718E+05 &      68.51 &     222.58 \\
 1089 & C & 6.93E+02 &    1.19323 &  6.3354E-09 &    20.0001 & 9.01E-06 & 1.704E+05 &      68.69 &     219.59 \\
 1090 & C & 6.70E+02 &    1.19596 &  6.3048E-09 &    20.0443 & 8.77E-06 & 1.690E+05 &      68.77 &     216.70 \\
 1091 & C & 6.47E+02 &    1.19806 &  6.2785E-09 &    20.0886 & 8.53E-06 & 1.676E+05 &      68.87 &     213.71 \\
 1092 & C & 6.24E+02 &    1.19947 &  6.2570E-09 &    20.1328 & 8.29E-06 & 1.661E+05 &      68.89 &     210.59 \\
 1093 & C & 6.01E+02 &    1.20015 &  6.2359E-09 &    20.1770 & 8.05E-06 & 1.641E+05 &      69.26 &     206.20 \\
 1094 & C & 5.62E+02 &    1.19937 &  6.2257E-09 &    20.2518 & 7.64E-06 & 1.622E+05 &      68.85 &     201.84 \\
 1095 & C & 5.43E+02 &    1.19803 &  6.2251E-09 &    20.2890 & 7.44E-06 & 1.609E+05 &      67.97 &     198.80 \\
 1096 & C & 5.23E+02 &    1.19464 &  6.2343E-09 &    20.3261 & 7.24E-06 & 1.589E+05 &      68.02 &     194.03 \\
 1097 & C & 4.85E+02 &    1.18825 &  6.2610E-09 &    20.3999 & 6.83E-06 & 1.566E+05 &      67.27 &     188.34 \\
 1098 & C & 4.59E+02 &    1.18033 &  6.2989E-09 &    20.4506 & 6.55E-06 & 1.546E+05 &      65.73 &     183.40 \\
 1099 & C & 4.33E+02 &    1.16997 &  6.3530E-09 &    20.5007 & 6.27E-06 & 1.525E+05 &      64.23 &     178.20 \\
 1100 & C & 4.07E+02 &    1.15723 &  6.4257E-09 &    20.5502 & 5.98E-06 & 1.504E+05 &      62.33 &     172.73 \\
 1101 & C & 3.82E+02 &    1.14352 &  6.5118E-09 &    20.5989 & 5.70E-06 & 1.483E+05 &      59.95 &     167.44 \\
 1102 & C & 3.61E+02 &    1.12949 &  6.6089E-09 &    20.6392 & 5.46E-06 & 1.464E+05 &      57.33 &     162.39 \\
 1103 & C & 3.41E+02 &    1.11461 &  6.7225E-09 &    20.6787 & 5.23E-06 & 1.444E+05 &      54.47 &     157.26 \\
 1104 & C & 3.21E+02 &    1.09927 &  6.8531E-09 &    20.7158 & 5.00E-06 & 1.424E+05 &      51.18 &     152.06 \\
 1105 & C & 3.02E+02 &    1.08331 &  7.0035E-09 &    20.7520 & 4.77E-06 & 1.404E+05 &      47.43 &     146.69 \\
 1106 & C & 2.83E+02 &    1.06508 &  7.1717E-09 &    20.7873 & 4.55E-06 & 1.382E+05 &      43.00 &     141.22 \\
 1107 & C & 2.65E+02 &    1.04501 &  7.3455E-09 &    20.8216 & 4.32E-06 & 1.360E+05 &      37.81 &     135.85 \\
 1108 & C & 2.48E+02 &    1.02372 &  7.5117E-09 &    20.8535 & 4.10E-06 & 1.338E+05 &      32.08 &     130.70 \\
 1109 & C & 2.31E+02 &    1.00600 &  7.6295E-09 &    20.8846 & 3.88E-06 & 1.319E+05 &      26.60 &     126.77 \\
 1110 & C & 2.22E+02 &    0.99256 &  7.7030E-09 &    20.9025 & 3.76E-06 & 1.306E+05 &      22.40 &     123.99 \\
 1111 & C & 2.13E+02 &    0.97901 &  7.7615E-09 &    20.9201 & 3.63E-06 & 1.292E+05 &      19.15 &     121.31 \\
 1112 & C & 2.03E+02 &    0.96557 &  7.8026E-09 &    20.9376 & 3.50E-06 & 1.277E+05 &      16.16 &     118.76 \\
 1113 & C & 1.94E+02 &    0.95303 &  7.8243E-09 &    20.9549 & 3.38E-06 & 1.263E+05 &      13.55 &     116.42 \\
 1114 & C & 1.86E+02 &    0.94136 &  7.8283E-09 &    20.9708 & 3.26E-06 & 1.250E+05 &      11.38 &     114.27 \\
 1115 & C & 1.78E+02 &    0.92946 &  7.8150E-09 &    20.9866 & 3.15E-06 & 1.236E+05 &       9.54 &     112.21 \\
 1116 & C & 1.70E+02 &    0.91738 &  7.7833E-09 &    21.0024 & 3.03E-06 & 1.222E+05 &       7.92 &     110.24 \\
 1117 & C & 1.63E+02 &    0.90353 &  7.7240E-09 &    21.0183 & 2.91E-06 & 1.206E+05 &       6.47 &     108.10 \\
 1118 & C & 1.53E+02 &    0.88804 &  7.6268E-09 &    21.0386 & 2.77E-06 & 1.187E+05 &       5.08 &     105.83 \\
 1119 & C & 1.44E+02 &    0.87366 &  7.5046E-09 &    21.0592 & 2.62E-06 & 1.169E+05 &       3.90 &     103.80 \\
 1120 & C & 1.36E+02 &    0.86054 &  7.3626E-09 &    21.0776 & 2.49E-06 & 1.152E+05 &       3.00 &     102.00 \\
 1121 & C & 1.28E+02 &    0.84800 &  7.1942E-09 &    21.0964 & 2.36E-06 & 1.134E+05 &       2.30 &     100.28 \\
 1122 & C & 1.20E+02 &    0.83628 &  6.9999E-09 &    21.1155 & 2.22E-06 & 1.115E+05 &       1.75 &      98.63 \\
 1123 & C & 1.12E+02 &    0.82528 &  6.7743E-09 &    21.1352 & 2.09E-06 & 1.096E+05 &       1.32 &      96.99 \\
 1124 & C & 1.04E+02 &    0.81748 &  6.5834E-09 &    21.1566 & 1.96E-06 & 1.080E+05 &       1.00 &      95.77 \\
 1125 & C & 1.00E+02 &    0.81258 &  6.4480E-09 &    21.1676 & 1.89E-06 & 1.069E+05 &       0.80 &      94.97 \\
 1126 & C & 9.62E+01 &    0.80607 &  6.2478E-09 &    21.1788 & 1.82E-06 & 1.054E+05 &       0.67 &      93.86 \\
 1127 & C & 8.92E+01 &    0.80011 &  6.0434E-09 &    21.1996 & 1.69E-06 & 1.038E+05 &       0.52 &      92.80 \\
 1128 & C & 8.57E+01 &    0.79626 &  5.9007E-09 &    21.2104 & 1.63E-06 & 1.028E+05 &       0.42 &      92.09 \\
 1129 & C & 8.22E+01 &    0.79196 &  5.7334E-09 &    21.2214 & 1.57E-06 & 1.016E+05 &       0.35 &      91.30 \\
 1130 & C & 7.79E+01 &    0.78715 &  5.5390E-09 &    21.2356 & 1.49E-06 & 1.002E+05 &       0.27 &      90.43 \\
 1131 & C & 7.36E+01 &    0.78269 &  5.3475E-09 &    21.2503 & 1.41E-06 & 9.880E+04 &       0.21 &      89.60 \\
 1132 & C & 6.98E+01 &    0.77884 &  5.1603E-09 &    21.2638 & 1.34E-06 & 9.747E+04 &       0.15 &      88.83 \\
 1133 & C & 6.61E+01 &    0.77531 &  4.9665E-09 &    21.2778 & 1.27E-06 & 9.609E+04 &       0.11 &      88.05 \\
 1134 & C & 6.23E+01 &    0.77266 &  4.8046E-09 &    21.2923 & 1.20E-06 & 9.494E+04 &       0.07 &      87.42 \\
 1135 & C & 6.00E+01 &    0.77075 &  4.6775E-09 &    21.3017 & 1.16E-06 & 9.403E+04 &       0.05 &      86.93 \\
 1136 & C & 5.77E+01 &    0.76804 &  4.4821E-09 &    21.3112 & 1.12E-06 & 9.261E+04 &       0.03 &      86.19 \\
 1137 & C & 5.32E+01 &    0.76529 &  4.2679E-09 &    21.3311 & 1.03E-06 & 9.105E+04 &       0.01 &      85.39 \\
 1138 & R & 5.04E+01 &    0.76325 &  4.1031E-09 &    21.3436 & 9.78E-07 & 8.983E+04 &       0.00 &      84.78 \\
 1139 & R & 4.77E+01 &    0.76129 &  3.9433E-09 &    21.3566 & 9.26E-07 & 8.863E+04 &       0.00 &      84.20 \\
 1140 & R & 4.53E+01 &    0.75959 &  3.7889E-09 &    21.3687 & 8.80E-07 & 8.746E+04 &       0.00 &      83.64 \\
 1141 & R & 4.29E+01 &    0.75813 &  3.6339E-09 &    21.3813 & 8.33E-07 & 8.627E+04 &       0.00 &      83.08 \\
 1142 & R & 4.06E+01 &    0.75688 &  3.4786E-09 &    21.3939 & 7.89E-07 & 8.505E+04 &       0.00 &      82.52 \\
 1143 & R & 3.83E+01 &    0.75581 &  3.3202E-09 &    21.4071 & 7.44E-07 & 8.378E+04 &       0.00 &      81.95 \\
 1144 & R & 3.60E+01 &    0.75487 &  3.1587E-09 &    21.4209 & 6.99E-07 & 8.246E+04 &       0.00 &      81.37 \\
 1145 & R & 3.37E+01 &    0.75414 &  3.0163E-09 &    21.4353 & 6.54E-07 & 8.126E+04 &       0.00 &      80.85 \\
 1146 & R & 3.20E+01 &    0.75353 &  2.8941E-09 &    21.4464 & 6.22E-07 & 8.021E+04 &       0.00 &      80.40 \\
 1147 & R & 3.04E+01 &    0.75290 &  2.7700E-09 &    21.4579 & 5.89E-07 & 7.913E+04 &       0.00 &      79.94 \\
 1148 & R & 2.87E+01 &    0.75225 &  2.6285E-09 &    21.4699 & 5.56E-07 & 7.786E+04 &       0.00 &      79.41 \\
 1149 & R & 2.67E+01 &    0.75167 &  2.4687E-09 &    21.4856 & 5.16E-07 & 7.638E+04 &       0.00 &      78.80 \\
 1150 & R & 2.46E+01 &    0.75132 &  2.3386E-09 &    21.5023 & 4.75E-07 & 7.514E+04 &       0.00 &      78.31 \\
 1151 & R & 2.34E+01 &    0.75114 &  2.2398E-09 &    21.5128 & 4.51E-07 & 7.417E+04 &       0.00 &      77.93 \\
 1152 & R & 2.22E+01 &    0.75102 &  2.1310E-09 &    21.5238 & 4.27E-07 & 7.308E+04 &       0.00 &      77.51 \\
 1153 & R & 2.08E+01 &    0.75100 &  2.0117E-09 &    21.5373 & 3.98E-07 & 7.185E+04 &       0.00 &      77.04 \\
 1154 & R & 1.93E+01 &    0.75107 &  1.9108E-09 &    21.5515 & 3.70E-07 & 7.077E+04 &       0.00 &      76.64 \\
 1155 & R & 1.84E+01 &    0.75118 &  1.8301E-09 &    21.5615 & 3.51E-07 & 6.989E+04 &       0.00 &      76.32 \\
 1156 & R & 1.75E+01 &    0.75135 &  1.7497E-09 &    21.5717 & 3.33E-07 & 6.899E+04 &       0.00 &      75.99 \\
 1157 & R & 1.65E+01 &    0.75159 &  1.6682E-09 &    21.5823 & 3.14E-07 & 6.805E+04 &       0.00 &      75.66 \\
 1158 & R & 1.56E+01 &    0.75193 &  1.5858E-09 &    21.5934 & 2.96E-07 & 6.707E+04 &       0.00 &      75.32 \\
 1159 & R & 1.47E+01 &    0.75232 &  1.5107E-09 &    21.6051 & 2.77E-07 & 6.616E+04 &       0.00 &      75.02 \\
 1160 & R & 1.39E+01 &    0.75277 &  1.4434E-09 &    21.6148 & 2.63E-07 & 6.531E+04 &       0.00 &      74.74 \\
 1161 & R & 1.32E+01 &    0.75333 &  1.3753E-09 &    21.6250 & 2.48E-07 & 6.444E+04 &       0.00 &      74.45 \\
 1162 & R & 1.25E+01 &    0.75402 &  1.3065E-09 &    21.6358 & 2.33E-07 & 6.352E+04 &       0.00 &      74.16 \\
 1163 & R & 1.17E+01 &    0.75502 &  1.2290E-09 &    21.6470 & 2.19E-07 & 6.246E+04 &       0.00 &      73.83 \\
 1164 & R & 1.09E+01 &    0.75652 &  1.1428E-09 &    21.6617 & 2.01E-07 & 6.122E+04 &       0.00 &      73.45 \\
 1165 & R & 9.97E+00 &    0.75795 &  1.0761E-09 &    21.6774 & 1.83E-07 & 6.023E+04 &       0.00 &      73.15 \\
 1166 & R & 9.49E+00 &    0.75910 &  1.0296E-09 &    21.6863 & 1.73E-07 & 5.950E+04 &       0.00 &      72.94 \\
 1167 & R & 9.01E+00 &    0.76037 &  9.8282E-10 &    21.6955 & 1.64E-07 & 5.876E+04 &       0.00 &      72.72 \\
 1168 & R & 8.54E+00 &    0.76193 &  9.3051E-10 &    21.7052 & 1.54E-07 & 5.789E+04 &       0.00 &      72.46 \\
 1169 & R & 7.96E+00 &    0.76379 &  8.7258E-10 &    21.7176 & 1.42E-07 & 5.689E+04 &       0.00 &      72.17 \\
 1170 & R & 7.38E+00 &    0.76562 &  8.1920E-10 &    21.7309 & 1.31E-07 & 5.593E+04 &       0.00 &      71.88 \\
 1171 & R & 6.89E+00 &    0.76732 &  7.7565E-10 &    21.7426 & 1.21E-07 & 5.511E+04 &       0.00 &      71.64 \\
 1172 & R & 6.51E+00 &    0.76905 &  7.3709E-10 &    21.7523 & 1.14E-07 & 5.435E+04 &       0.00 &      71.41 \\
 1173 & R & 6.12E+00 &    0.77070 &  7.0359E-10 &    21.7626 & 1.06E-07 & 5.367E+04 &       0.00 &      71.21 \\
 1174 & R & 5.84E+00 &    0.77221 &  6.7523E-10 &    21.7704 & 1.00E-07 & 5.308E+04 &       0.00 &      71.03 \\
 1175 & R & 5.56E+00 &    0.77414 &  6.4136E-10 &    21.7785 & 9.48E-08 & 5.234E+04 &       0.00 &      70.80 \\
 1176 & R & 5.17E+00 &    0.77624 &  6.0673E-10 &    21.7904 & 8.71E-08 & 5.156E+04 &       0.00 &      70.55 \\
 1177 & R & 4.87E+00 &    0.77813 &  5.7676E-10 &    21.7999 & 8.13E-08 & 5.085E+04 &       0.00 &      70.32 \\
 1178 & R & 4.58E+00 &    0.77975 &  5.5149E-10 &    21.8098 & 7.54E-08 & 5.024E+04 &       0.00 &      70.11 \\
 1179 & R & 4.37E+00 &    0.78112 &  5.3095E-10 &    21.8169 & 7.15E-08 & 4.972E+04 &       0.00 &      69.94 \\
 1180 & R & 4.17E+00 &    0.78266 &  5.1037E-10 &    21.8243 & 6.75E-08 & 4.918E+04 &       0.00 &      69.75 \\
 1181 & R & 3.96E+00 &    0.78445 &  4.8907E-10 &    21.8320 & 6.35E-08 & 4.861E+04 &       0.00 &      69.55 \\
 1182 & R & 3.74E+00 &    0.78647 &  4.6708E-10 &    21.8405 & 5.93E-08 & 4.799E+04 &       0.00 &      69.32 \\
 1183 & R & 3.53E+00 &    0.78864 &  4.4509E-10 &    21.8494 & 5.51E-08 & 4.735E+04 &       0.00 &      69.07 \\
 1184 & R & 3.31E+00 &    0.79092 &  4.2311E-10 &    21.8587 & 5.08E-08 & 4.669E+04 &       0.00 &      68.80 \\
 1185 & R & 3.09E+00 &    0.79285 &  4.0457E-10 &    21.8686 & 4.66E-08 & 4.610E+04 &       0.00 &      68.55 \\
 1186 & R & 2.94E+00 &    0.79437 &  3.8946E-10 &    21.8756 & 4.37E-08 & 4.560E+04 &       0.00 &      68.33 \\
 1187 & R & 2.78E+00 &    0.79580 &  3.7432E-10 &    21.8830 & 4.08E-08 & 4.509E+04 &       0.00 &      68.10 \\
 1188 & R & 2.63E+00 &    0.79704 &  3.5915E-10 &    21.8906 & 3.79E-08 & 4.456E+04 &       0.00 &      67.85 \\
 1189 & R & 2.48E+00 &    0.79808 &  3.4600E-10 &    21.8985 & 3.49E-08 & 4.409E+04 &       0.00 &      67.62 \\
 1190 & R & 2.37E+00 &    0.79898 &  3.3489E-10 &    21.9045 & 3.28E-08 & 4.367E+04 &       0.00 &      67.42 \\
 1191 & R & 2.26E+00 &    0.80002 &  3.2165E-10 &    21.9107 & 3.07E-08 & 4.317E+04 &       0.00 &      67.16 \\
 1192 & R & 2.10E+00 &    0.80087 &  3.0983E-10 &    21.9195 & 2.78E-08 & 4.271E+04 &       0.00 &      66.93 \\
 1193 & R & 2.02E+00 &    0.80139 &  3.0157E-10 &    21.9244 & 2.62E-08 & 4.238E+04 &       0.00 &      66.76 \\
 1194 & R & 1.94E+00 &    0.80182 &  2.9326E-10 &    21.9295 & 2.47E-08 & 4.204E+04 &       0.00 &      66.58 \\
 1195 & R & 1.86E+00 &    0.80213 &  2.8491E-10 &    21.9346 & 2.31E-08 & 4.169E+04 &       0.00 &      66.40 \\
 1196 & R & 1.78E+00 &    0.80227 &  2.7606E-10 &    21.9400 & 2.16E-08 & 4.131E+04 &       0.00 &      66.20 \\
 1197 & R & 1.69E+00 &    0.80215 &  2.6667E-10 &    21.9460 & 1.99E-08 & 4.091E+04 &       0.00 &      65.99 \\
 1198 & R & 1.59E+00 &    0.80180 &  2.5928E-10 &    21.9523 & 1.81E-08 & 4.058E+04 &       0.00 &      65.83 \\
 1199 & R & 1.54E+00 &    0.80112 &  2.5125E-10 &    21.9559 & 1.72E-08 & 4.023E+04 &       0.00 &      65.65 \\
 1200 & R & 1.44E+00 &    0.80003 &  2.4310E-10 &    21.9634 & 1.53E-08 & 3.986E+04 &       0.00 &      65.48 \\
 1201 & R & 1.39E+00 &    0.79900 &  2.3729E-10 &    21.9672 & 1.43E-08 & 3.960E+04 &       0.00 &      65.36 \\
 1202 & R & 1.34E+00 &    0.79775 &  2.3130E-10 &    21.9715 & 1.32E-08 & 3.933E+04 &       0.00 &      65.25 \\
 1203 & R & 1.29E+00 &    0.79556 &  2.2262E-10 &    21.9757 & 1.22E-08 & 3.894E+04 &       0.00 &      65.10 \\
 1204 & R & 1.19E+00 &    0.79285 &  2.1391E-10 &    21.9841 & 1.03E-08 & 3.854E+04 &       0.00 &      64.98 \\
 1205 & R & 1.14E+00 &    0.78970 &  2.0556E-10 &    21.9883 & 9.40E-09 & 3.817E+04 &       0.00 &      64.90 \\
 1206 & R & 1.05E+00 &    0.78603 &  1.9736E-10 &    21.9964 & 7.68E-09 & 3.780E+04 &       0.00 &      64.86 \\
 1207 & R & 1.01E+00 &    0.78257 &  1.9068E-10 &    22.0003 & 6.88E-09 & 3.751E+04 &       0.00 &      64.86 \\
 1208 & R & 9.50E-01 &    0.77882 &  1.8424E-10 &    22.0064 & 5.69E-09 & 3.723E+04 &       0.00 &      64.89 \\
 1209 & R & 9.17E-01 &    0.77546 &  1.7903E-10 &    22.0098 & 5.03E-09 & 3.700E+04 &       0.00 &      64.95 \\
 1210 & R & 8.76E-01 &    0.77201 &  1.7411E-10 &    22.0141 & 4.24E-09 & 3.680E+04 &       0.00 &      65.03 \\
 1211 & R & 8.49E-01 &    0.76795 &  1.6877E-10 &    22.0172 & 3.69E-09 & 3.658E+04 &       0.00 &      65.15 \\
 1212 & R & 8.06E-01 &    0.76388 &  1.6380E-10 &    22.0221 & 2.83E-09 & 3.637E+04 &       0.00 &      65.30 \\
 1213 & R & 7.85E-01 &    0.76046 &  1.5988E-10 &    22.0245 & 2.42E-09 & 3.622E+04 &       0.00 &      65.44 \\
 1214 & R & 7.57E-01 &    0.75691 &  1.5601E-10 &    22.0280 & 1.85E-09 & 3.606E+04 &       0.00 &      65.60 \\
 1215 & R & 7.39E-01 &    0.75246 &  1.5144E-10 &    22.0303 & 1.47E-09 & 3.589E+04 &       0.00 &      65.83 \\
 1216 & R & 7.04E-01 &    0.74810 &  1.4719E-10 &    22.0348 & 7.64E-10 & 3.573E+04 &       0.00 &      66.07 \\
 1217 & R & 6.91E-01 &    0.74443 &  1.4378E-10 &    22.0366 & 4.96E-10 & 3.560E+04 &       0.00 &      66.29 \\
 1218 & R & 6.67E-01 &    0.74192 &  1.4152E-10 &    22.0399 & 0.00E+00 & 3.552E+04 &       0.00 &      66.45 \\

\hline
\end{longtable}

\clearpage
\begin{longtable}{cccccccccc}

\caption[table]{Outermost 1 $\rso$ of a 20 $\mso$ model at solar metallicity ($Z$=0.02). The table shows the physical variables at t=\ttwenty years. Columns 
contain the progressive grid point of the model, the status (R=radiative, C=convective), the optical depth $\tau$, the opacity $\varkappa$,
the density $\rho$, the radius R, the value $M_{*}-M_r$ (where $M_*$ is the total stellar mass and $M_r$ is the mass coordinate), the temperature T,
the convective velocity $v_c$ and the local sound speed $c_s$. All the values are in cgs units if not otherwise specified. 
} \label{table20} \\

\hline\hline
\\[-2ex]
  
 Grid point &  STAT &  $\tau$ &  $\varkappa$ &  $\rho$ &    $R$ [$\rso$] &      $M_{*}-M_r$ [g] &  T[K] &       $v_c$ [km~s$^{-1}$] &  $c_s$ [km~s$^{-1}$] \\
\\[-2ex]
\hline
\endfirsthead

\\[-2ex]
\hline\hline
 Grid point &  STAT &  $\tau$ &  $\varkappa$ &  $\rho$&    $R$ [$\rso$] &      $M_{*}-M_r$ [g] &  T[K] &       $v_c$ [km~s$^{-1}$] &  $c_s$ [km~s$^{-1}$] \\
\\[-2ex]
\hline

\endhead

  \multicolumn{3}{l}{{Continued on Next Page\ldots}} \\
\endfoot

\\[-2ex]
\hline 
\endlastfoot

\hline

 1009 & R & 1.88E+04 &    1.15006 &  9.1148E-07 &     9.4516 & 3.91E-05 & 3.415E+05 &       0.00 &     110.14 \\
 1010 & R & 1.80E+04 &    1.16843 &  8.4671E-07 &     9.4634 & 3.70E-05 & 3.363E+05 &       0.00 &     109.93 \\
 1011 & R & 1.68E+04 &    1.18658 &  7.8819E-07 &     9.4826 & 3.39E-05 & 3.313E+05 &       0.00 &     109.74 \\
 1012 & R & 1.61E+04 &    1.19963 &  7.4912E-07 &     9.4929 & 3.23E-05 & 3.279E+05 &       0.00 &     109.61 \\
 1013 & R & 1.55E+04 &    1.21788 &  6.9847E-07 &     9.5037 & 3.08E-05 & 3.232E+05 &       0.00 &     109.45 \\
 1014 & R & 1.45E+04 &    1.24209 &  6.3760E-07 &     9.5220 & 2.83E-05 & 3.172E+05 &       0.00 &     109.26 \\
 1015 & R & 1.35E+04 &    1.26170 &  5.9299E-07 &     9.5413 & 2.59E-05 & 3.125E+05 &       0.00 &     109.14 \\
 1016 & R & 1.30E+04 &    1.27584 &  5.6324E-07 &     9.5516 & 2.47E-05 & 3.093E+05 &       0.00 &     109.06 \\
 1017 & R & 1.24E+04 &    1.29061 &  5.3416E-07 &     9.5624 & 2.36E-05 & 3.060E+05 &       0.00 &     108.98 \\
 1018 & R & 1.19E+04 &    1.30554 &  5.0665E-07 &     9.5733 & 2.24E-05 & 3.027E+05 &       0.00 &     108.92 \\
 1019 & R & 1.15E+04 &    1.32091 &  4.8003E-07 &     9.5841 & 2.14E-05 & 2.994E+05 &       0.00 &     108.86 \\
 1020 & R & 1.10E+04 &    1.33683 &  4.5439E-07 &     9.5954 & 2.03E-05 & 2.961E+05 &       0.00 &     108.81 \\
 1021 & R & 1.05E+04 &    1.35328 &  4.2968E-07 &     9.6065 & 1.93E-05 & 2.928E+05 &       0.00 &     108.77 \\
 1022 & R & 1.01E+04 &    1.37094 &  4.0503E-07 &     9.6182 & 1.83E-05 & 2.893E+05 &       0.00 &     108.73 \\
 1023 & R & 9.62E+03 &    1.38833 &  3.8248E-07 &     9.6306 & 1.73E-05 & 2.860E+05 &       0.00 &     108.71 \\
 1024 & R & 9.24E+03 &    1.40532 &  3.6200E-07 &     9.6416 & 1.65E-05 & 2.828E+05 &       0.00 &     108.69 \\
 1025 & R & 8.85E+03 &    1.42367 &  3.4159E-07 &     9.6531 & 1.57E-05 & 2.795E+05 &       0.00 &     108.68 \\
 1026 & R & 8.45E+03 &    1.44389 &  3.2125E-07 &     9.6653 & 1.48E-05 & 2.761E+05 &       0.00 &     108.67 \\
 1027 & R & 8.05E+03 &    1.46417 &  3.0276E-07 &     9.6783 & 1.40E-05 & 2.728E+05 &       0.00 &     108.67 \\
 1028 & R & 7.72E+03 &    1.48419 &  2.8611E-07 &     9.6896 & 1.33E-05 & 2.697E+05 &       0.00 &     108.68 \\
 1029 & R & 7.38E+03 &    1.50595 &  2.6954E-07 &     9.7016 & 1.26E-05 & 2.664E+05 &       0.00 &     108.68 \\
 1030 & R & 7.04E+03 &    1.52815 &  2.5408E-07 &     9.7142 & 1.19E-05 & 2.632E+05 &       0.00 &     108.68 \\
 1031 & R & 6.73E+03 &    1.55060 &  2.3969E-07 &     9.7260 & 1.13E-05 & 2.601E+05 &       0.00 &     108.68 \\
 1032 & R & 6.42E+03 &    1.57310 &  2.2639E-07 &     9.7385 & 1.07E-05 & 2.570E+05 &       0.00 &     108.67 \\
 1033 & R & 6.15E+03 &    1.59542 &  2.1415E-07 &     9.7499 & 1.02E-05 & 2.541E+05 &       0.00 &     108.66 \\
 1034 & R & 5.88E+03 &    1.62281 &  2.0030E-07 &     9.7619 & 9.69E-06 & 2.505E+05 &       0.00 &     108.63 \\
 1035 & R & 5.53E+03 &    1.65803 &  1.8493E-07 &     9.7784 & 9.02E-06 & 2.463E+05 &       0.00 &     108.57 \\
 1036 & R & 5.17E+03 &    1.69089 &  1.7276E-07 &     9.7961 & 8.35E-06 & 2.426E+05 &       0.00 &     108.48 \\
 1037 & R & 4.94E+03 &    1.72391 &  1.6211E-07 &     9.8077 & 7.94E-06 & 2.392E+05 &       0.00 &     108.36 \\
 1038 & R & 4.64E+03 &    1.76662 &  1.5016E-07 &     9.8242 & 7.39E-06 & 2.350E+05 &       0.00 &     108.14 \\
 1039 & R & 4.32E+03 &    1.80536 &  1.4067E-07 &     9.8420 & 6.84E-06 & 2.314E+05 &       0.00 &     107.88 \\
 1040 & R & 4.12E+03 &    1.83771 &  1.3352E-07 &     9.8539 & 6.50E-06 & 2.284E+05 &       0.00 &     107.60 \\
 1041 & R & 3.91E+03 &    1.86954 &  1.2702E-07 &     9.8664 & 6.15E-06 & 2.255E+05 &       0.00 &     107.28 \\
 1042 & R & 3.73E+03 &    1.90044 &  1.2115E-07 &     9.8776 & 5.85E-06 & 2.226E+05 &       0.00 &     106.90 \\
 1043 & R & 3.55E+03 &    1.93471 &  1.1542E-07 &     9.8894 & 5.56E-06 & 2.197E+05 &       0.00 &     106.44 \\
 1044 & R & 3.37E+03 &    1.97305 &  1.0986E-07 &     9.9017 & 5.26E-06 & 2.165E+05 &       0.00 &     105.86 \\
 1045 & R & 3.18E+03 &    2.01222 &  1.0492E-07 &     9.9146 & 4.96E-06 & 2.134E+05 &       0.00 &     105.21 \\
 1046 & R & 3.01E+03 &    2.05131 &  1.0056E-07 &     9.9260 & 4.71E-06 & 2.105E+05 &       0.00 &     104.47 \\
 1047 & C & 2.85E+03 &    2.09321 &  9.6390E-08 &     9.9379 & 4.46E-06 & 2.073E+05 &       0.27 &     103.59 \\
 1048 & C & 2.68E+03 &    2.13709 &  9.2418E-08 &     9.9502 & 4.21E-06 & 2.039E+05 &       3.48 &     102.53 \\
 1049 & C & 2.50E+03 &    2.17706 &  8.8976E-08 &     9.9630 & 3.96E-06 & 2.007E+05 &       5.74 &     101.40 \\
 1050 & C & 2.36E+03 &    2.21180 &  8.6007E-08 &     9.9739 & 3.76E-06 & 1.976E+05 &       7.30 &     100.24 \\
 1051 & C & 2.21E+03 &    2.24604 &  8.3153E-08 &     9.9851 & 3.55E-06 & 1.943E+05 &       8.51 &      98.93 \\
 1052 & C & 2.07E+03 &    2.27681 &  8.0595E-08 &     9.9967 & 3.35E-06 & 1.911E+05 &       9.43 &      97.55 \\
 1053 & C & 1.94E+03 &    2.30324 &  7.8309E-08 &    10.0070 & 3.17E-06 & 1.879E+05 &      10.10 &      96.14 \\
 1054 & C & 1.80E+03 &    2.32459 &  7.6279E-08 &    10.0177 & 2.99E-06 & 1.848E+05 &      10.53 &      94.69 \\
 1055 & C & 1.69E+03 &    2.34028 &  7.4483E-08 &    10.0269 & 2.84E-06 & 1.818E+05 &      10.74 &      93.24 \\
 1056 & C & 1.58E+03 &    2.35064 &  7.2749E-08 &    10.0363 & 2.69E-06 & 1.786E+05 &      10.81 &      91.67 \\
 1057 & C & 1.47E+03 &    2.35334 &  7.1321E-08 &    10.0460 & 2.54E-06 & 1.758E+05 &      10.61 &      90.24 \\
 1058 & C & 1.39E+03 &    2.34958 &  7.0171E-08 &    10.0528 & 2.43E-06 & 1.734E+05 &      10.26 &      89.00 \\
 1059 & C & 1.31E+03 &    2.33995 &  6.9027E-08 &    10.0598 & 2.33E-06 & 1.708E+05 &       9.87 &      87.70 \\
 1060 & C & 1.23E+03 &    2.32417 &  6.7878E-08 &    10.0668 & 2.22E-06 & 1.682E+05 &       9.34 &      86.34 \\
 1061 & C & 1.16E+03 &    2.30414 &  6.6802E-08 &    10.0740 & 2.12E-06 & 1.657E+05 &       8.69 &      85.05 \\
 1062 & C & 1.09E+03 &    2.28134 &  6.5791E-08 &    10.0801 & 2.03E-06 & 1.633E+05 &       8.00 &      83.83 \\
 1063 & C & 1.03E+03 &    2.25516 &  6.4766E-08 &    10.0863 & 1.94E-06 & 1.610E+05 &       7.32 &      82.61 \\
 1064 & C & 9.71E+02 &    2.22633 &  6.3719E-08 &    10.0923 & 1.86E-06 & 1.585E+05 &       6.62 &      81.39 \\
 1065 & C & 9.12E+02 &    2.19204 &  6.2614E-08 &    10.0984 & 1.77E-06 & 1.561E+05 &       5.91 &      80.16 \\
 1066 & C & 8.55E+02 &    2.15002 &  6.1433E-08 &    10.1046 & 1.69E-06 & 1.535E+05 &       5.16 &      78.90 \\
 1067 & C & 8.00E+02 &    2.10271 &  6.0183E-08 &    10.1109 & 1.60E-06 & 1.509E+05 &       4.39 &      77.68 \\
 1068 & C & 7.48E+02 &    2.05208 &  5.8853E-08 &    10.1170 & 1.52E-06 & 1.483E+05 &       3.66 &      76.48 \\
 1069 & C & 6.97E+02 &    2.01013 &  5.7714E-08 &    10.1233 & 1.44E-06 & 1.462E+05 &       3.03 &      75.53 \\
 1070 & C & 6.68E+02 &    1.97885 &  5.6821E-08 &    10.1270 & 1.40E-06 & 1.447E+05 &       2.59 &      74.84 \\
 1071 & C & 6.40E+02 &    1.94778 &  5.5882E-08 &    10.1308 & 1.35E-06 & 1.431E+05 &       2.26 &      74.14 \\
 1072 & C & 6.11E+02 &    1.91749 &  5.4897E-08 &    10.1346 & 1.30E-06 & 1.415E+05 &       1.97 &      73.45 \\
 1073 & C & 5.84E+02 &    1.88866 &  5.3907E-08 &    10.1386 & 1.25E-06 & 1.399E+05 &       1.71 &      72.79 \\
 1074 & C & 5.59E+02 &    1.86001 &  5.2917E-08 &    10.1422 & 1.21E-06 & 1.383E+05 &       1.49 &      72.15 \\
 1075 & C & 5.34E+02 &    1.83056 &  5.1882E-08 &    10.1459 & 1.17E-06 & 1.368E+05 &       1.28 &      71.52 \\
 1076 & C & 5.10E+02 &    1.80057 &  5.0800E-08 &    10.1497 & 1.13E-06 & 1.352E+05 &       1.09 &      70.88 \\
 1077 & C & 4.86E+02 &    1.76619 &  4.9509E-08 &    10.1536 & 1.08E-06 & 1.333E+05 &       0.91 &      70.17 \\
 1078 & C & 4.57E+02 &    1.72789 &  4.7985E-08 &    10.1586 & 1.03E-06 & 1.312E+05 &       0.73 &      69.36 \\
 1079 & C & 4.29E+02 &    1.69264 &  4.6470E-08 &    10.1638 & 9.72E-07 & 1.292E+05 &       0.56 &      68.61 \\
 1080 & C & 4.04E+02 &    1.66094 &  4.4977E-08 &    10.1686 & 9.24E-07 & 1.272E+05 &       0.43 &      67.90 \\
 1081 & C & 3.80E+02 &    1.63122 &  4.3415E-08 &    10.1735 & 8.75E-07 & 1.252E+05 &       0.32 &      67.18 \\
 1082 & C & 3.56E+02 &    1.60288 &  4.1784E-08 &    10.1786 & 8.26E-07 & 1.232E+05 &       0.23 &      66.46 \\
 1083 & C & 3.33E+02 &    1.57488 &  4.0035E-08 &    10.1838 & 7.78E-07 & 1.210E+05 &       0.16 &      65.72 \\
 1084 & C & 3.09E+02 &    1.54737 &  3.8160E-08 &    10.1896 & 7.26E-07 & 1.186E+05 &       0.10 &      64.94 \\
 1085 & C & 2.86E+02 &    1.52243 &  3.6302E-08 &    10.1957 & 6.75E-07 & 1.163E+05 &       0.05 &      64.18 \\
 1086 & C & 2.65E+02 &    1.49992 &  3.4470E-08 &    10.2015 & 6.29E-07 & 1.140E+05 &       0.01 &      63.44 \\
 1087 & R & 2.45E+02 &    1.48240 &  3.2925E-08 &    10.2075 & 5.83E-07 & 1.121E+05 &       0.00 &      62.83 \\
 1088 & R & 2.32E+02 &    1.46981 &  3.1697E-08 &    10.2115 & 5.54E-07 & 1.105E+05 &       0.00 &      62.35 \\
 1089 & R & 2.19E+02 &    1.45895 &  3.0510E-08 &    10.2156 & 5.24E-07 & 1.090E+05 &       0.00 &      61.88 \\
 1090 & R & 2.08E+02 &    1.44955 &  2.9369E-08 &    10.2194 & 4.99E-07 & 1.075E+05 &       0.00 &      61.43 \\
 1091 & R & 1.97E+02 &    1.44085 &  2.8205E-08 &    10.2234 & 4.73E-07 & 1.060E+05 &       0.00 &      60.97 \\
 1092 & R & 1.86E+02 &    1.43173 &  2.6869E-08 &    10.2275 & 4.47E-07 & 1.042E+05 &       0.00 &      60.43 \\
 1093 & R & 1.72E+02 &    1.42204 &  2.5352E-08 &    10.2329 & 4.15E-07 & 1.022E+05 &       0.00 &      59.81 \\
 1094 & R & 1.59E+02 &    1.41419 &  2.4109E-08 &    10.2385 & 3.82E-07 & 1.005E+05 &       0.00 &      59.29 \\
 1095 & R & 1.50E+02 &    1.40808 &  2.3160E-08 &    10.2421 & 3.63E-07 & 9.911E+04 &       0.00 &      58.89 \\
 1096 & R & 1.42E+02 &    1.40244 &  2.2244E-08 &    10.2458 & 3.44E-07 & 9.779E+04 &       0.00 &      58.50 \\
 1097 & R & 1.35E+02 &    1.39730 &  2.1365E-08 &    10.2493 & 3.27E-07 & 9.650E+04 &       0.00 &      58.12 \\
 1098 & R & 1.28E+02 &    1.39246 &  2.0485E-08 &    10.2529 & 3.10E-07 & 9.519E+04 &       0.00 &      57.73 \\
 1099 & R & 1.21E+02 &    1.38794 &  1.9605E-08 &    10.2565 & 2.93E-07 & 9.385E+04 &       0.00 &      57.34 \\
 1100 & R & 1.14E+02 &    1.38372 &  1.8710E-08 &    10.2603 & 2.76E-07 & 9.246E+04 &       0.00 &      56.93 \\
 1101 & R & 1.08E+02 &    1.37985 &  1.7798E-08 &    10.2642 & 2.60E-07 & 9.101E+04 &       0.00 &      56.51 \\
 1102 & R & 1.01E+02 &    1.37684 &  1.6995E-08 &    10.2684 & 2.43E-07 & 8.970E+04 &       0.00 &      56.13 \\
 1103 & R & 9.59E+01 &    1.37462 &  1.6306E-08 &    10.2716 & 2.31E-07 & 8.855E+04 &       0.00 &      55.80 \\
 1104 & R & 9.10E+01 &    1.37291 &  1.5607E-08 &    10.2749 & 2.19E-07 & 8.736E+04 &       0.00 &      55.46 \\
 1105 & R & 8.61E+01 &    1.37164 &  1.4811E-08 &    10.2783 & 2.07E-07 & 8.597E+04 &       0.00 &      55.07 \\
 1106 & R & 8.01E+01 &    1.37106 &  1.3914E-08 &    10.2829 & 1.92E-07 & 8.435E+04 &       0.00 &      54.60 \\
 1107 & R & 7.41E+01 &    1.37119 &  1.3184E-08 &    10.2877 & 1.77E-07 & 8.299E+04 &       0.00 &      54.22 \\
 1108 & R & 7.04E+01 &    1.37161 &  1.2631E-08 &    10.2907 & 1.68E-07 & 8.192E+04 &       0.00 &      53.91 \\
 1109 & R & 6.68E+01 &    1.37234 &  1.2022E-08 &    10.2938 & 1.59E-07 & 8.072E+04 &       0.00 &      53.57 \\
 1110 & R & 6.26E+01 &    1.37335 &  1.1354E-08 &    10.2977 & 1.48E-07 & 7.936E+04 &       0.00 &      53.18 \\
 1111 & R & 5.83E+01 &    1.37444 &  1.0789E-08 &    10.3018 & 1.37E-07 & 7.817E+04 &       0.00 &      52.85 \\
 1112 & R & 5.55E+01 &    1.37559 &  1.0338E-08 &    10.3047 & 1.30E-07 & 7.720E+04 &       0.00 &      52.57 \\
 1113 & R & 5.27E+01 &    1.37700 &  9.8872E-09 &    10.3076 & 1.24E-07 & 7.620E+04 &       0.00 &      52.29 \\
 1114 & R & 5.00E+01 &    1.37874 &  9.4310E-09 &    10.3107 & 1.17E-07 & 7.516E+04 &       0.00 &      51.99 \\
 1115 & R & 4.72E+01 &    1.38088 &  8.9689E-09 &    10.3139 & 1.10E-07 & 7.407E+04 &       0.00 &      51.69 \\
 1116 & R & 4.44E+01 &    1.38319 &  8.5480E-09 &    10.3173 & 1.03E-07 & 7.305E+04 &       0.00 &      51.40 \\
 1117 & R & 4.22E+01 &    1.38562 &  8.1697E-09 &    10.3201 & 9.76E-08 & 7.211E+04 &       0.00 &      51.13 \\
 1118 & R & 4.00E+01 &    1.38845 &  7.7869E-09 &    10.3230 & 9.22E-08 & 7.112E+04 &       0.00 &      50.85 \\
 1119 & R & 3.78E+01 &    1.39177 &  7.3995E-09 &    10.3261 & 8.67E-08 & 7.010E+04 &       0.00 &      50.56 \\
 1120 & R & 3.56E+01 &    1.39617 &  6.9626E-09 &    10.3293 & 8.12E-08 & 6.890E+04 &       0.00 &      50.23 \\
 1121 & R & 3.29E+01 &    1.40204 &  6.4740E-09 &    10.3336 & 7.45E-08 & 6.750E+04 &       0.00 &      49.83 \\
 1122 & R & 3.02E+01 &    1.40739 &  6.0948E-09 &    10.3381 & 6.79E-08 & 6.636E+04 &       0.00 &      49.51 \\
 1123 & R & 2.88E+01 &    1.41162 &  5.8291E-09 &    10.3406 & 6.43E-08 & 6.554E+04 &       0.00 &      49.27 \\
 1124 & R & 2.73E+01 &    1.41634 &  5.5609E-09 &    10.3433 & 6.08E-08 & 6.468E+04 &       0.00 &      49.03 \\
 1125 & R & 2.59E+01 &    1.42226 &  5.2595E-09 &    10.3461 & 5.72E-08 & 6.368E+04 &       0.00 &      48.74 \\
 1126 & R & 2.41E+01 &    1.42985 &  4.9241E-09 &    10.3497 & 5.29E-08 & 6.252E+04 &       0.00 &      48.41 \\
 1127 & R & 2.23E+01 &    1.43837 &  4.6133E-09 &    10.3536 & 4.86E-08 & 6.140E+04 &       0.00 &      48.09 \\
 1128 & R & 2.08E+01 &    1.44656 &  4.3587E-09 &    10.3570 & 4.50E-08 & 6.043E+04 &       0.00 &      47.81 \\
 1129 & R & 1.96E+01 &    1.45480 &  4.1322E-09 &    10.3598 & 4.22E-08 & 5.954E+04 &       0.00 &      47.54 \\
 1130 & R & 1.84E+01 &    1.46274 &  3.9346E-09 &    10.3628 & 3.93E-08 & 5.873E+04 &       0.00 &      47.30 \\
 1131 & R & 1.76E+01 &    1.47005 &  3.7666E-09 &    10.3651 & 3.73E-08 & 5.802E+04 &       0.00 &      47.08 \\
 1132 & R & 1.67E+01 &    1.47948 &  3.5652E-09 &    10.3675 & 3.52E-08 & 5.712E+04 &       0.00 &      46.80 \\
 1133 & R & 1.55E+01 &    1.48989 &  3.3581E-09 &    10.3710 & 3.24E-08 & 5.617E+04 &       0.00 &      46.50 \\
 1134 & R & 1.45E+01 &    1.50018 &  3.1779E-09 &    10.3738 & 3.02E-08 & 5.529E+04 &       0.00 &      46.22 \\
 1135 & R & 1.36E+01 &    1.51024 &  3.0255E-09 &    10.3767 & 2.80E-08 & 5.452E+04 &       0.00 &      45.96 \\
 1136 & R & 1.29E+01 &    1.51938 &  2.9012E-09 &    10.3789 & 2.66E-08 & 5.386E+04 &       0.00 &      45.74 \\
 1137 & R & 1.23E+01 &    1.52942 &  2.7765E-09 &    10.3811 & 2.51E-08 & 5.318E+04 &       0.00 &      45.51 \\
 1138 & R & 1.16E+01 &    1.54073 &  2.6471E-09 &    10.3834 & 2.36E-08 & 5.244E+04 &       0.00 &      45.24 \\
 1139 & R & 1.09E+01 &    1.55335 &  2.5130E-09 &    10.3860 & 2.20E-08 & 5.164E+04 &       0.00 &      44.94 \\
 1140 & R & 1.02E+01 &    1.56677 &  2.3784E-09 &    10.3887 & 2.05E-08 & 5.079E+04 &       0.00 &      44.61 \\
 1141 & R & 9.50E+00 &    1.58081 &  2.2433E-09 &    10.3916 & 1.89E-08 & 4.988E+04 &       0.00 &      44.25 \\
 1142 & R & 8.78E+00 &    1.59453 &  2.1289E-09 &    10.3946 & 1.73E-08 & 4.908E+04 &       0.00 &      43.91 \\
 1143 & R & 8.28E+00 &    1.60744 &  2.0356E-09 &    10.3968 & 1.62E-08 & 4.838E+04 &       0.00 &      43.60 \\
 1144 & R & 7.77E+00 &    1.62179 &  1.9422E-09 &    10.3991 & 1.51E-08 & 4.765E+04 &       0.00 &      43.25 \\
 1145 & R & 7.26E+00 &    1.63733 &  1.8489E-09 &    10.4016 & 1.41E-08 & 4.688E+04 &       0.00 &      42.87 \\
 1146 & R & 6.75E+00 &    1.65130 &  1.7685E-09 &    10.4041 & 1.30E-08 & 4.618E+04 &       0.00 &      42.50 \\
 1147 & R & 6.37E+00 &    1.66300 &  1.7008E-09 &    10.4060 & 1.22E-08 & 4.555E+04 &       0.00 &      42.15 \\
 1148 & R & 5.99E+00 &    1.67593 &  1.6205E-09 &    10.4080 & 1.14E-08 & 4.477E+04 &       0.00 &      41.69 \\
 1149 & C & 5.46E+00 &    1.68633 &  1.5492E-09 &    10.4109 & 1.03E-08 & 4.404E+04 &       0.07 &      41.23 \\
 1150 & C & 5.18E+00 &    1.69348 &  1.4995E-09 &    10.4125 & 9.75E-09 & 4.350E+04 &       0.07 &      40.87 \\
 1151 & C & 4.89E+00 &    1.70032 &  1.4498E-09 &    10.4142 & 9.17E-09 & 4.294E+04 &       0.07 &      40.47 \\
 1152 & C & 4.61E+00 &    1.70654 &  1.4000E-09 &    10.4159 & 8.59E-09 & 4.236E+04 &       0.08 &      40.03 \\
 1153 & C & 4.32E+00 &    1.71194 &  1.3475E-09 &    10.4177 & 8.02E-09 & 4.171E+04 &       0.08 &      39.52 \\
 1154 & C & 4.00E+00 &    1.71555 &  1.2921E-09 &    10.4198 & 7.38E-09 & 4.100E+04 &       0.08 &      38.92 \\
 1155 & C & 3.68E+00 &    1.71615 &  1.2488E-09 &    10.4219 & 6.74E-09 & 4.042E+04 &       0.09 &      38.41 \\
 1156 & C & 3.51E+00 &    1.71368 &  1.2020E-09 &    10.4231 & 6.38E-09 & 3.977E+04 &       0.09 &      37.80 \\
 1157 & C & 3.16E+00 &    1.70637 &  1.1546E-09 &    10.4257 & 5.67E-09 & 3.909E+04 &       0.09 &      37.14 \\
 1158 & C & 2.98E+00 &    1.69773 &  1.1208E-09 &    10.4270 & 5.31E-09 & 3.859E+04 &       0.10 &      36.66 \\
 1159 & C & 2.79E+00 &    1.68575 &  1.0858E-09 &    10.4285 & 4.92E-09 & 3.806E+04 &       0.09 &      36.16 \\
 1160 & C & 2.61E+00 &    1.67097 &  1.0512E-09 &    10.4300 & 4.55E-09 & 3.754E+04 &       0.13 &      35.67 \\
 1161 & C & 2.43E+00 &    1.65364 &  1.0168E-09 &    10.4315 & 4.18E-09 & 3.701E+04 &       0.12 &      35.21 \\
 1162 & C & 2.27E+00 &    1.63409 &  9.8245E-10 &    10.4330 & 3.84E-09 & 3.649E+04 &       0.11 &      34.79 \\
 1163 & C & 2.11E+00 &    1.61186 &  9.4679E-10 &    10.4345 & 3.49E-09 & 3.596E+04 &       0.13 &      34.41 \\
 1164 & C & 1.95E+00 &    1.59444 &  9.2036E-10 &    10.4361 & 3.15E-09 & 3.556E+04 &       0.14 &      34.16 \\
 1165 & C & 1.88E+00 &    1.58330 &  9.0408E-10 &    10.4367 & 3.00E-09 & 3.532E+04 &       0.16 &      34.02 \\
 1166 & C & 1.81E+00 &    1.56475 &  8.7896E-10 &    10.4374 & 2.85E-09 & 3.496E+04 &       0.16 &      33.84 \\
 1167 & C & 1.68E+00 &    1.54394 &  8.5291E-10 &    10.4389 & 2.56E-09 & 3.458E+04 &       0.16 &      33.68 \\
 1168 & C & 1.61E+00 &    1.52883 &  8.3498E-10 &    10.4396 & 2.41E-09 & 3.433E+04 &       0.17 &      33.59 \\
 1169 & C & 1.55E+00 &    1.51274 &  8.1655E-10 &    10.4404 & 2.26E-09 & 3.408E+04 &       0.19 &      33.52 \\
 1170 & C & 1.48E+00 &    1.48977 &  7.9113E-10 &    10.4412 & 2.11E-09 & 3.373E+04 &       0.19 &      33.43 \\
 1171 & R & 1.38E+00 &    1.46832 &  7.6805E-10 &    10.4425 & 1.87E-09 & 3.342E+04 &       0.00 &      33.38 \\
 1172 & R & 1.34E+00 &    1.44895 &  7.4758E-10 &    10.4431 & 1.77E-09 & 3.316E+04 &       0.00 &      33.34 \\
 1173 & R & 1.25E+00 &    1.42850 &  7.2624E-10 &    10.4442 & 1.57E-09 & 3.289E+04 &       0.00 &      33.32 \\
 1174 & R & 1.21E+00 &    1.41409 &  7.1132E-10 &    10.4448 & 1.47E-09 & 3.270E+04 &       0.00 &      33.31 \\
 1175 & R & 1.17E+00 &    1.39428 &  6.9089E-10 &    10.4454 & 1.37E-09 & 3.245E+04 &       0.00 &      33.31 \\
 1176 & R & 1.10E+00 &    1.36990 &  6.6582E-10 &    10.4465 & 1.20E-09 & 3.215E+04 &       0.00 &      33.32 \\
 1177 & R & 1.05E+00 &    1.34597 &  6.4121E-10 &    10.4474 & 1.05E-09 & 3.187E+04 &       0.00 &      33.35 \\
 1178 & R & 9.87E-01 &    1.32442 &  6.1900E-10 &    10.4485 & 8.97E-10 & 3.162E+04 &       0.00 &      33.38 \\
 1179 & R & 9.44E-01 &    1.30661 &  6.0086E-10 &    10.4493 & 7.84E-10 & 3.141E+04 &       0.00 &      33.42 \\
 1180 & R & 9.07E-01 &    1.28575 &  5.8007E-10 &    10.4500 & 6.86E-10 & 3.119E+04 &       0.00 &      33.47 \\
 1181 & R & 8.57E-01 &    1.26672 &  5.6146E-10 &    10.4510 & 5.48E-10 & 3.099E+04 &       0.00 &      33.52 \\
 1182 & R & 8.32E-01 &    1.25106 &  5.4637E-10 &    10.4515 & 4.81E-10 & 3.083E+04 &       0.00 &      33.57 \\
 1183 & R & 7.98E-01 &    1.23145 &  5.2773E-10 &    10.4522 & 3.85E-10 & 3.064E+04 &       0.00 &      33.64 \\
 1184 & R & 7.63E-01 &    1.20815 &  5.0587E-10 &    10.4531 & 2.84E-10 & 3.042E+04 &       0.00 &      33.74 \\
 1185 & R & 7.21E-01 &    1.18955 &  4.8863E-10 &    10.4541 & 1.60E-10 & 3.025E+04 &       0.00 &      33.84 \\
 1186 & R & 7.04E-01 &    1.17182 &  4.7235E-10 &    10.4545 & 1.11E-10 & 3.010E+04 &       0.00 &      33.93 \\
 1187 & R & 6.67E-01 &    1.15923 &  4.6089E-10 &    10.4556 & 0.00E+00 & 2.999E+04 &       0.00 &      34.01 \\

\hline
\end{longtable}
\end{center}

\end{appendix}
\end{document}